\documentclass[10pt]{article}

\usepackage[backend=biber,
            sorting=none,
            maxnames=99,
            style=nature]{biblatex}
\usepackage[margin=1in]{geometry}
\usepackage[english]{babel}
\usepackage{pdfpages}
\usepackage[font=small]{caption}
\usepackage{float}
\usepackage{amsmath}
\usepackage[bookmarks=false]{hyperref}
\usepackage[capitalise]{cleveref}
\usepackage{authblk}
\usepackage{csquotes}
\usepackage{subfiles}

\defbibfilter{onlymain}{%
  segment=0
}

\defbibfilter{onlyappendix}{%
  not segment=0
}

\newcommand{\extendeddata}{%
    \renewcommand{\figurename}{Extended Data Figure}%
    \captionsetup[figure]{labelformat=simple,labelsep=period}%
    \setcounter{figure}{0}
}

\newcommand{\suppmat}{%
    \renewcommand{\figurename}{Supplementary Figure}%
    \renewcommand{\tablename}{Supplementary Figure}
    \captionsetup[figure]{labelformat=simple,labelsep=period}%
    \setcounter{figure}{0}
}

\newcommand{\resetfigurelabel}{%
  \renewcommand{\figurename}{Figure}%
  \captionsetup[figure]{labelformat=default,labelsep=colon}%
}

\title{\Large Extending intraday solar forecast horizons with deep generative models}
\author[a, b]{A. Carpentieri\thanks{Corresponding author: alberto.carpentieri@bfh.ch}}
\author[a]{D. Folini}
\author[c]{J. Leinonen\thanks{Currently at NVIDIA Corporation}}
\author[b, d]{A. Meyer}

\affil[a]{\footnotesize{Institute for Atmospheric and Climate Science, ETH Zurich, Universitaetstrasse 16, 8092 Zurich, Switzerland}}
\affil[b]{\footnotesize{School of Engineering and Computer Science, Bern University of Applied Sciences, Quellgasse 21, 2501 Biel, Switzerland}}
\affil[c]{\footnotesize{Federal Office of Meteorology and Climatology MeteoSwiss, Via ai Monti 146, 6605 Locarno-Monti, Switzerland}}
\affil[d]{\footnotesize{Department of Geoscience and Remote Sensing, TU Delft, Stevinweg 1, 2628 CN Delft, Netherlands}}
\date{}

\begin{document}

\maketitle

\section*{Abstract}
Surface solar irradiance (SSI) plays a crucial role in tackling climate change – as an abundant, non-fossil energy source, exploited primarily via photovoltaic (PV) energy production. With the growing contribution of SSI to total energy production, the stability of the latter is challenged by the intermittent character of the former, arising primarily from cloud effects. Mitigating this stability challenge requires accurate, uncertainty-aware, near real-time, regional-scale SSI forecasts with lead times of minutes to a few hours, enabling robust real-time energy grid management. State-of-the-art nowcasting methods typically meet only some of these requirements. Here we present SHADECast, a deep generative diffusion model for the probabilistic spatiotemporal nowcasting of SSI, conditioned on deterministic aspects of cloud evolution to guide the probabilistic ensemble forecast, and based on near real-time satellite data. We demonstrate that SHADECast provides improved forecast quality, reliability, and accuracy in different weather scenarios. Our model produces realistic and spatiotemporally consistent predictions outperforming the state of the art by 15\% in the continuous ranked probability score (CRPS) over different regions up to 512 km $\times$ 512 km with lead times of 15-120 min. Conditioning the ensemble generation on deterministic forecasts improves reliability and performance by more than 7\% on CRPS. Our approach empowers grid operators and energy traders to make informed decisions, ensuring stability and facilitating the seamless integration of PV energy across multiple locations simultaneously.

\pagebreak

\section*{Main}
\label{main_chapter}
Harvesting solar energy resources is an essential pillar in efforts to mitigate climate change\cite{Yang_review}. Photovoltaic (PV) power generation increased by 26\% on 2022, accounting for two-thirds of the increase in global renewable capacity for 2023 \cite{IEA2023}. In concert with the growing relevance of PV for total energy production, the challenge arising from the intermittent character of surface solar irradiance (SSI) increases. Power production and consumption, linked via transmission and storage capacities, should be closely balanced at any moment in time. The naturally arising volatility of PV production, primarily due to changing cloudiness, impacts the reliability of the electricity grid \cite{SMITH2022}. A key element in dealing with this challenge - and the topic of this paper - are regional-scale, near real-time, uncertainty-aware SSI forecasts with lead times of minutes to hours. Such forecasts enable strategic planning of energy production from alternate sources, such as gas turbines \cite{Pv_integration, IEA2023}, facilitate the proactive scheduling of energy-intensive industrial operations \cite{machines10090730, Yang_review}, and thus reduce operation uncertainty and stand-by costs \cite{Haupt2019, Yang_review}. 

The relevance of the topic spurred progress in SSI forecasting. Yet, there remains ample room and an urgent need for substantial further improvement. State-of-the-art methods span a wide range of approaches. For short lead times of up to a few hours, which are the focus of this work, data-driven methods prevail with numerical weather prediction \cite{wang_nwp_datadriven} playing only a minor role. One distinguishing feature is the input data used. Ground-based in-situ measurements of SSI have the advantage of being highly accurate, but their limited spatial representativeness \cite{Representativeness} discourages their exclusive use for regional-scale forecasts.
 
Satellite-derived solar irradiance estimates offer a trade-off between accuracy and spatial coverage, which makes them highly suitable for short-term SSI forecasting over extended regions, enabling simultaneous SSI forecasts for multiple sites \cite{PALETTA2023}. Various data-driven SSI forecast methods that rely on satellite data exist, notably statistical methods \cite{HAMMER1999, Urbich2018, AYET_Analog_2018, WANG2019, aicardi2022, CARPENTIERI2023}, deep learning models \cite{knol_2021_convrnn_vs_of, nielsen_irradiancenet_2021, GALLO2022, son_lstmgan_2023, reg_det_gan}, and hybrid approaches that also incorporate numerical weather predictions \cite{ARBIZUBARRENA2017, hatanaka_diffusion_2023}. 

The majority of these approaches, despite using satellite data as input, provide forecasts only at individual locations \cite{PALETTA2023, reg_det_gan}, which is not suitable for managing arbitrarily large grids \cite{reg_det_gan}. Approaches providing regional scale forecasts are mostly deterministic \cite{Zhang_2022_SSR_Int_Est}, which again limits their practical use for lack of forecast uncertainty quantification. Also, existing deterministic models tend to generate blurry forecasts, as illustrated by recent studies comparing convolutional recurrent neural networks and optical flow methods \cite{knol_2021_convrnn_vs_of}. The blurriness results from the mean squared error (MSE) minimization, which causes predictions to converge towards the mean of the distribution of all possible future SSI evolutions \cite{babaeizadeh2018stochastic, Ravuri2021}. The resulting forecasts lack the spatial granularity required to accurately represent the stochastic spatiotemporal behaviour of SSI.  

Spatiotemporal regional scale SSI forecasts with uncertainty quantification are still scarce. In \cite{AYET_Analog_2018}, an Analog Ensemble method is applied to retrieve past SSI field sequences (analogs) based on four similarity metrics and project them into the future to generate an ensemble of forecasts. The analog-based approach can be effective but requires a huge amount of past data and a complete search in the dataset for each forecast. A more flexible ensemble-based approach is proposed in \cite{CARPENTIERI2023}, where scale-dependent autoregressive (AR) models are applied to probabilistically forecast cloudiness fields in a Monte Carlo sampling approach. However, linear AR models assume stationarity in the data, making the model unable to predict distribution shifts. Differently, in \cite{nielsen_irradiancenet_2021}, a deterministic convolutional long short-term memory (ConvLSTM) model is modified to directly forecast the probability of each pixel value inside different ranges. This classification based procedure drastically increases the dimensionality of the output by a factor of 240, making it impractical for large-area and multi-step forecasts. 

Here we present the Solar High-resolution Adaptive Diffusion Ensemble forecasting model (SHADECast), producing uncertainty-aware regional-scale SSI forecasts that model probabilistic cloud formation, evolution, and dissipation, conditioned on a data-driven deterministic cloud field forecast. Our approach is novel in that it combines insight from atmospheric physics -  leading us to split the task into a deterministic part upon which a probabilistic part then acts - with inspiration from probabilistic video forecasting, where generative deep learning models have emerged as the new state of the art due to their adeptness in modeling data distributions, enabling the sampling of realistic future scenarios \cite{Yang_video_LDM_2023}. Notably, diffusion models \cite{Sohl-Dickstein2015, Ho2020} have exhibited superior performance in image and video generation tasks \cite{dhariwal_LDM_vs_GANs_2021, Yang_video_LDM_2023}. In precipitation nowcasting, they provide superior characterization of the distribution of possible outcomes compared to generative adversarial networks \cite{leinonen2023, Ravuri2021}.

SHADECast is, to the best of our knowledge, the first uncertainty-aware, physics-inspired deterministic-probabilistic, satellite-based regional-scale forecast model for intraday SSI forecasts.  As we are going to demonstrate, SHADECast produces skillful, sharp and reliable, realistic solar forecasts without blurring under variable weather conditions, thanks also to our innovative, physics motivated splitting of the task at hand. 

We assess our model's performance by comparing it with three benchmark models. Two benchmark models are probabilistic: SolarSTEPS \cite{CARPENTIERI2023_SS}, which was shown to outperform several benchmark SSI forecasting models, and an adaptation of the precipitation nowcasting model, LDCast \cite{leinonen2023}, trained to forecast cloudiness fields. A deterministic model (ConvLSTM \cite{nielsen_irradiancenet_2021}) is also employed as benchmark to highlight the benefits of probabilistic modeling.  
Our model outperforms state-of-the-art models by improving on key performance metrics, such as the CRPS, by 15\%. A 120-minute SSI ensemble forecast of SHADECast is, on average, as skillful as a 94-minute SSI forecast of the state-of-the-art probabilistic SSI forecasting ensemble-based model, SolarSTEPS \cite{CARPENTIERI2023_SS}. 

\section*{Surface solar irradiance}

SSI can be expressed as the product of the clear-sky SSI, $\textrm{SSI}_{\textrm{cs}}$, and the clear-sky index, CSI, so $\textrm{SSI} = \textrm{CSI}\cdot\textrm{SSI}_{\textrm{cs}}$. The clear-sky SSI is an estimate of SSI in the absence of clouds. $\textrm{SSI}_{\textrm{cs}}$ mainly depends on the solar zenith angle (SZA), its diurnal and annual cycle, and to a minor degree on aerosols and atmospheric trace gases like water vapor. The remaining most relevant factor affecting SSI are clouds, which are also the most difficult component to forecast. CSI is a dimensionless variable that quantifies the degree of cloudiness, which makes CSI a particularly suitable variable to forecast \cite{nielsen_irradiancenet_2021, CARPENTIERI2023_SS}. SHADECast forecasts spatial cloudiness fields expressed in terms of CSI, based on satellite-derived CSI estimates for lead times of up to 2 hours.

The temporal evolution of cloudiness fields may be seen as a composite of wind-driven cloud advection and cloud evolution - the formation, growth, and dissipation of clouds - governed by processes such as microphysics and turbulence \cite{CARPENTIERI2023_SS}. While cloud advection and cloud evolution cannot be separated from each other in a strict physical sense, it pays off to do so in the context of forecasting, as we demonstrate below. SHADECast invokes a probabilistic method for cloud evolution, guided by a deterministic forecast of the wind-advected cloud field.

\begin{figure}[h]
    \centering
    \includegraphics[width=\textwidth, trim=0cm 0cm 1cm 0cm, clip]{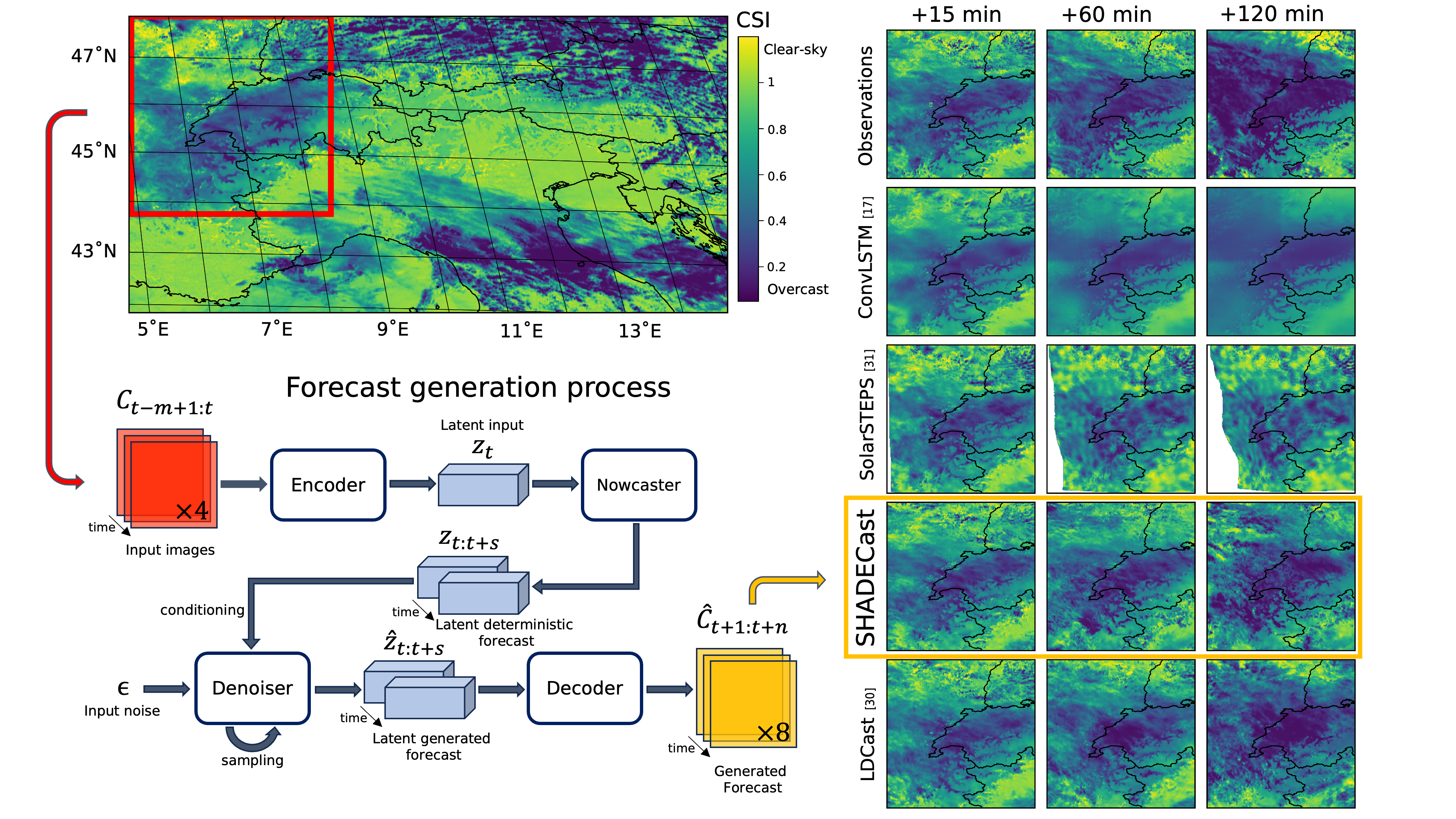}    
    \caption{\textbf{Upper left panel}: Example CSI field of 24 Feb. 2016 at 11:45 UTC. The red box highlights the region forecasted in the right panel.
    \textbf{Lower left panel}: SHADECast forecast generation pipeline. The input CSI fields ($C_{t-m+1:t}$) are fed to the encoder, which projects the image sequence to the latent space, obtaining $z_t$. Then, the deterministic nowcaster forecasts the future latent representation of the CSI fields ($z_{t+1:t+s}$), where $s$ is the lead time in the latent space, which can differ from $n$ due to data compression. The latent forecast is, then, fed to the denoiser together with Gaussian noise $\epsilon$. The pseudo linear multi-step (PLMS) sampler employs the denoiser to generate an ensemble member. The decoder finally decompresses the latent ensemble forecast, obtaining $\hat{C}_{t+1:t+n}$.  
    \textbf{Right panel}: Forecasts made by SHADECast (yellow box) and benchmark models for lead times up to 120 minutes. For SHADECast, LDCast and SolarSTEPS the ensemble member chosen is the one with the lowest average root mean squared error (RMSE). The first row shows the satellite-derived CSI fields.} 
    \label{GenerationProcess}
\end{figure}

\section*{Generative short-term forecasting}
Our goal is to generate an ensemble forecast consisting of future CSI fields $\hat{C}$ that are consistent with CSI fields $C$ observed shortly before the time when the forecast is made. Based on a sequence of $m$ observed fields $C_{t-m+1:t}$, we want to forecast $n$ future fields $\hat{C}_{t+1:t+n}$ by means of a forecasting process $f_{\theta}$ starting at time $t$,
\begin{equation}
    \hat{C}_{t+1:t+n} = f_{\theta} \bigl( C_{t-m+1:t}, \epsilon \bigr)
    \label{Eq2}
\end{equation}
with free parameters $\theta$ whose optimal values $\theta^{*}$ are determined by minimizing the distance between the estimated conditional probability distribution of forecasted cloudiness fields $p_{\theta}(\hat{C}_{t+1:t+n}|C_{t-m+1:t})$ and the actual distribution of the future fields $p(C_{t+1:t+n}|C_{t-m+1:t})$. The normally distributed random variable  $\epsilon$ is sampled multiple times to draw individual ensemble members of the forecast from $p_{\theta}$ according to \Cref{Eq2}. SHADECast offers a concrete realization of this general concept.

The SHADECast forecast generation pipeline, depicted in \Cref{GenerationProcess}, integrates a variational autoencoder (VAE) for data compression, a latent deterministic nowcaster based on Adaptive Fourier Neural Operator (AFNO) blocks \cite{guibas_afno_2022, pathak2022}, and a latent diffusion model represented by the denoiser. These components collaboratively forecast an ensemble of future cloudiness field sequences. The nowcaster's deterministic forecast guides the ensemble generation by the denoiser \cite{Ho2020}. With respect to previous SSI nowcasting methods and to LDCast, an important conceptual innovation of our model lies in the decomposition of the forecasting task into a deterministic forecast (nowcaster) for large-scale dynamics and a probabilistic ensemble generation (diffusion) to model high-uncertainty regions.

The encoder, nowcaster, and denoiser are trained independently. The training data comprises seven years of satellite data over central Europe with $768\times384$ pixels in total (see \Cref{GenerationProcess}). To economize on memory usage, training is done on sequences of $128\times128$ pixel satellite images. Once trained, the model generates a 2-hour forecast for a $256\times256$ pixel region (red box in \Cref{GenerationProcess}) in less than 7 seconds on an Nvidia T100 GPU. The evaluation is conducted using forecast ensembles with 10 members on three different regions (see Extended Data Figure \ref{Patches}).

\section*{Clouds forming, evolving, dissipating}
\begin{figure}[h!]
    \centering
    \includegraphics[width=\textwidth]{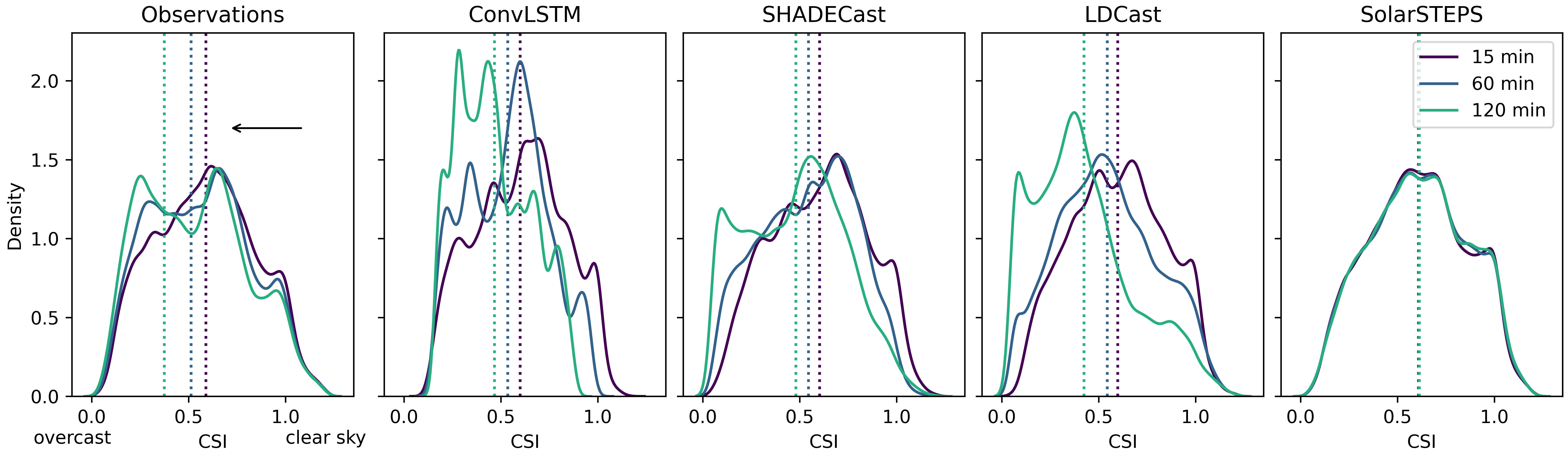} 
    \caption{The estimated probability density distributions of the CSI pixel values relative to the case study presented in \Cref{GenerationProcess}. 
    The probability distributions are shown for the ground truth satellite-derived CSI fields (Observations), for SHADECast and three benchmark models. For SHADECast, LDCast and SolarSTEPS, the chosen ensemble member is the best performing one in terms of RMSE. The dotted vertical lines represent the distribution mean.} 
    \label{Test1_KDE}
\end{figure}

In \Cref{GenerationProcess}, we show an example of a forecast generated by SHADECast. We present the ground truth in the first row, a deterministic forecast generated by a convolutional LSTM model based on \cite{nielsen_irradiancenet_2021} in the second row, SHADECast in the fourth row and the two benchmark models in the remaining rows. This particular case study is selected to exemplify the dynamic nature of cloud evolution throughout the forecast period. This phenomenon is visually represented by observing the shift in the mean of the CSI distribution towards lower values, as illustrated in \Cref{Test1_KDE}.

The ConvLSTM forecast is relatively accurate within the initial 15 minutes, but its quality gradually diminishes afterwards due to increasing blurriness and the inability of the deterministic model to handle uncertainty. The observed lack of small-scale structures is linked to the convergence towards the mean \cite{babaeizadeh2018stochastic} due to the pixel-level MSE minimisation performed in the training. As highlighted in the introduction, our objective is modeling the distribution of potential outcomes, as the average of all outcomes (MSE minimum) does not necessarily align with the most probable outcome. SHADECast effectively simulates diverse cloudiness evolution in high-uncertainty regions, providing insights into variations that might appear indistinct in deterministic forecasts. On the other hand, the model can recognize low-uncertainty regions and keep them relatively unaltered among the ensemble members. In \Cref{GenerationProcess}, the Alps region (bottom right area in the map) remains cloud-free throughout the 2-hour period. Similar patterns in the same region are evident in the SHADECast ensemble members but not in the benchmark probabilistic models (LDCast and SolarSTEPS). This case study demonstrates the adaptability of SHADECast in capturing ground truth uncertainty and projecting it into the forecast ensemble while retaining the less uncertain patterns. Additional forecast examples for the three test regions are also presented (see Extended Data Figures \ref{Test1_Forex}, \ref{Test3_Forex}, \ref{Test5_Forex}).

A distinguishing feature of SHADECast is that it allows for changes of the CSI field probability density distribution over time, as shown in \Cref{Test1_KDE}. A scene can get more or less cloudy with time. This is a clear asset as compared to SolarSTEPS, which is limited by its underlying linear AR model to forecast stationary time series. This leads SolarSTEPS to produce fields that have approximately the same CSI distribution as the input, making it incapable of predicting scenarios where the weather situation drastically changes. This limitation is clearly visible in \Cref{GenerationProcess}, where the cloudy region expands significantly during the forecasted period, and even more so in \Cref{Test1_KDE}, which illustrates the distributions of CSI values for individual fields at three lead times. Also apparent is the narrowing of the distribution in the case of ConvLSTM, consistent with the overall tendency of this deterministic forecast to dump the tails of the CSI distribution in favor of mean values. This effect drastically reduces the accuracy in predicting extreme CSI values. On the other hand, SHADECast accurately follows the observed distributional shift and outperforms the benchmark models in predicting extreme values (see Supplementary Figure \ref{FSS}).

\section*{Performance evaluation}
\begin{figure}[h]
    \centering
    \includegraphics[width=\textwidth]{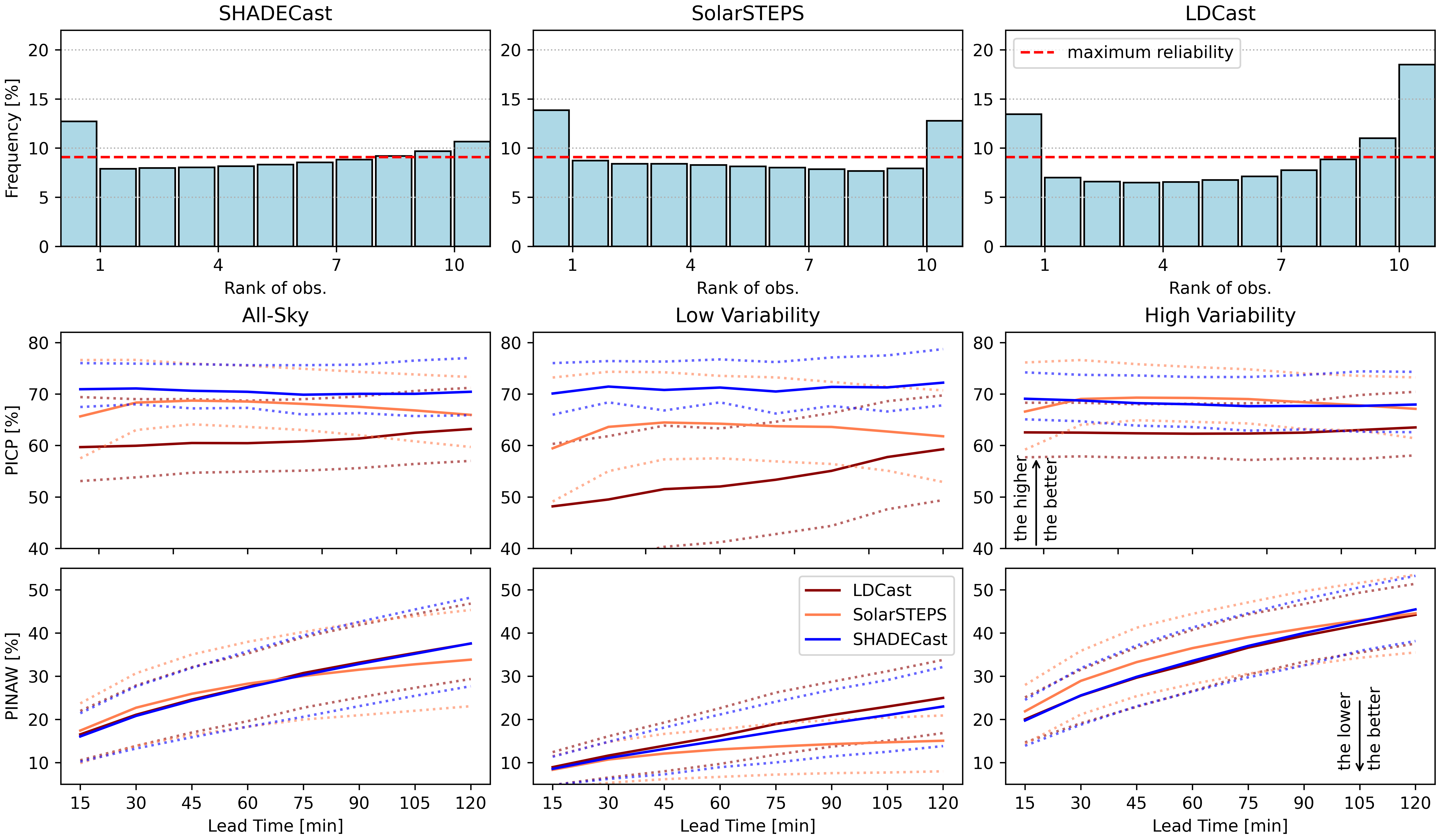} 
    \caption{\textbf{Upper panel}: rank histogram for the 10 ensemble members (x-axis), comprising the entire test set. Our model (SHADECast) clearly provides more reliable forecasts - frequencies closer to the maximum reliability line -. The high external columns on LDCast and SolarSTEPS rank histograms highlight the models overconfidence as more than 30\% of CSI values fall outside the ensemble forecasts. \textbf{Lower panel}: PICP and PINAW metrics computed on the test set across different lead times. The first measures the reliability (number of ground truth pixels falling inside the prediction interval), while the second measures the sharpness of the forecast (normalized width of the prediction interval). Both metrics are measured using a confidence interval of 90\%. The dotted lines represent the $25^{th}$ and $75^{th}$ percentile of the correspondent metric values over the entire test set.} 
    \label{Rank_Histograms}
\end{figure}

Common measures to evaluate ensemble forecast performance include (see Methods for further details) rank histograms, prediction interval coverage probability (PICP) and prediction interval normalized average width (PINAW), as well as the continuous ranked probability score (CRPS). Rank histograms shown in \Cref{Rank_Histograms} demonstrate that SHADECast produces significantly more reliable probabilistic forecasts compared to the benchmark models. One can notice the tendency of LDCast and SolarSTEPS to generate ensembles that tend to be overconfident, underestimating the uncertainty of cloudiness evolution. LDCast overestimates, in particular, the occurrence of overcast situations (low CSI). On the other hand, SHADECast can better model the uncertainty, providing significantly more reliable ensembles. The rank histograms are computed on the test set across three different regions (see \Cref{Patches}). In Supplementary Figure \ref{All_Rank_Histograms}, we provide the rank histograms for the three test regions, individually. The reliability of the models does not depend on the considered location.

Model reliability can also be quantified via the PICP, shown in the second row of \Cref{Rank_Histograms}, and the PINAW, also presented in \Cref{Rank_Histograms}. The first metric calculates the average number of pixels that fall within the ensemble prediction interval, with its width determined by the second metric. The average PICP is $\approx 70\%$ for ShADECast compared to $65\%$ and $60\%$ of SolarSTEPS and LDCast, respectively. The major improvement of SHADECast over LDCast can be noticed in the low-variability samples, where the model provides sharper predictions (lower PINAW) and achieves higher PICP. Instead, SolarSTEPS generally provides ensembles with lower variance, consequently achieving a lower PICP. 

\begin{figure}[h]
    \centering
    \includegraphics[width=\textwidth]{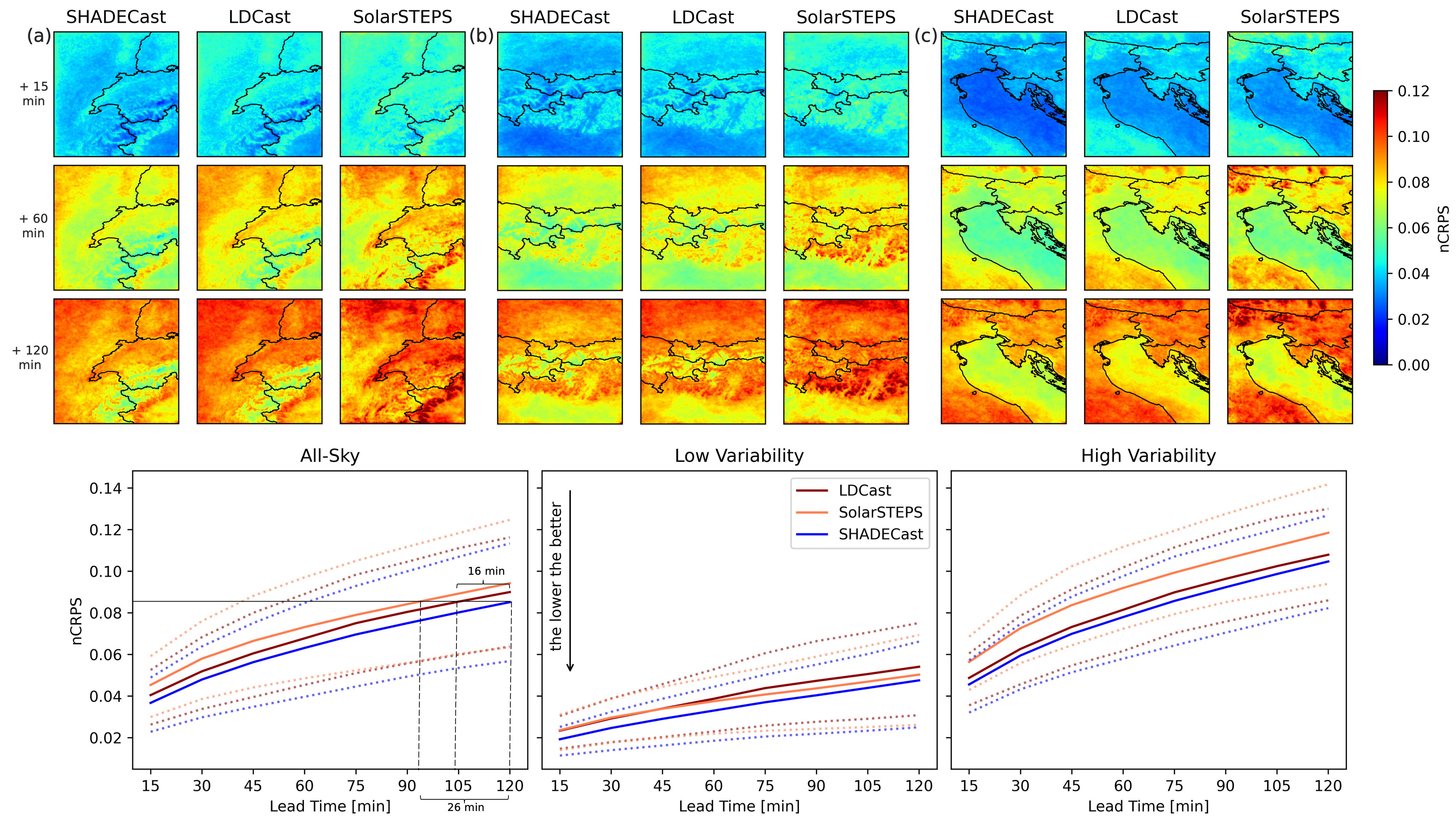}
    \caption{\textbf{Upper panel}: normalized Continuous Ranked Probability Score (nCRPS) averaged over the entire test set for the three test patches. The metric is shown for three lead times (+15, +60, +120 min) for SHADECast and the benchmark models. 
    \textbf{Lower panel}: Average, $25^{th}$ and $75^{th}$ percentiles of nCRPS are shown for the 8 lead times and for the three models. The metric is computed for all the forecasts in the test set for every pixel and then averaged. The solid lines represent the mean value for every lead time, while the dotted lines represent the percentiles. The values shown are averaged for the entire test set (All-sky) and for two subsets, representative of low-variability and high-variability cloudiness situations.} 
    \label{CRPS}
\end{figure}

\begin{figure}[H]
    \centering
    \includegraphics[width=\textwidth, trim = 0cm 4cm 0cm 4cm, clip]{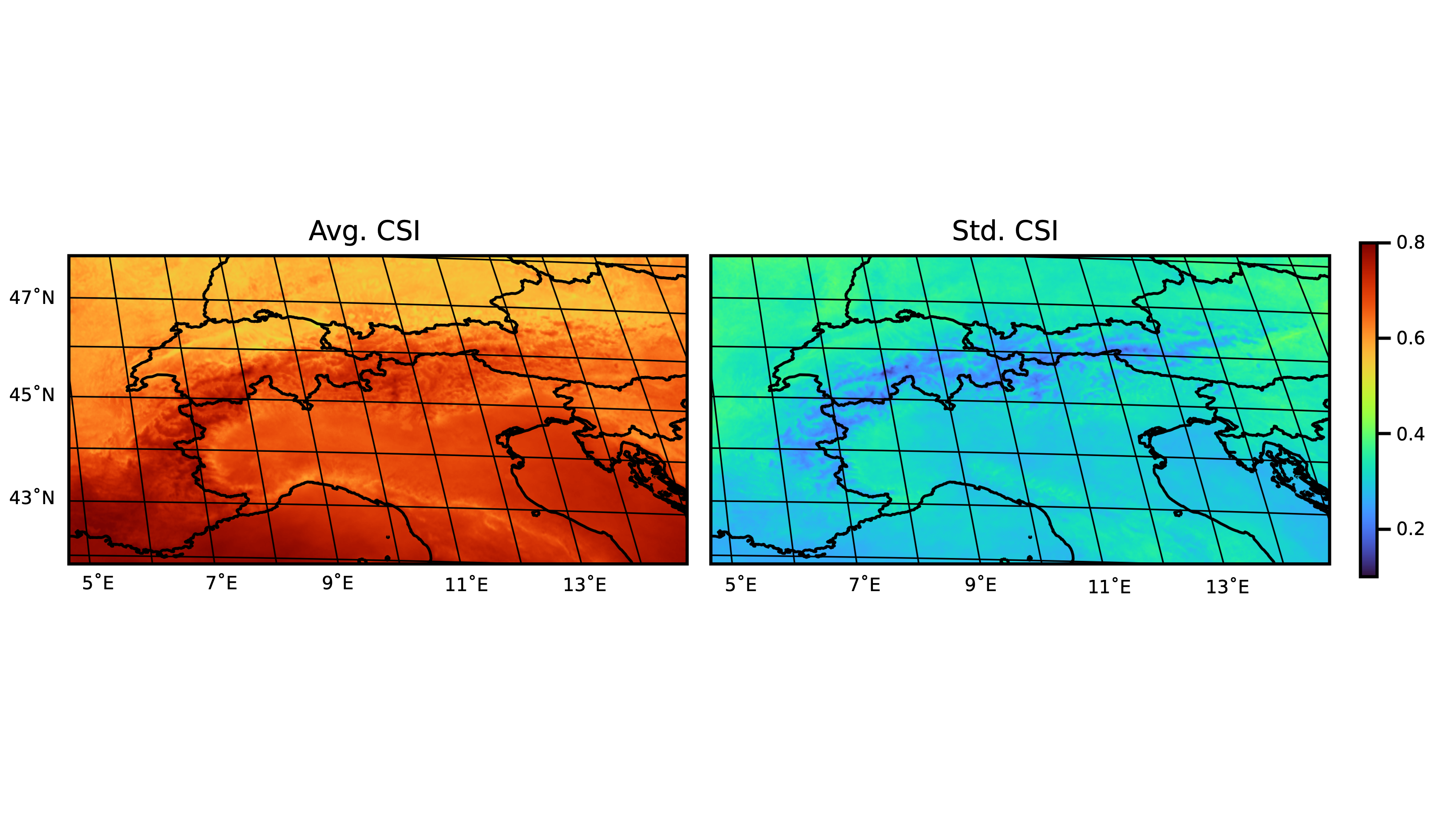} 
    \caption{Average and standard deviation at pixel level of CSI values computed on 500 hundred days sampled from the training set. \textbf{Left panel}: average CSI values for the HelioMont covered region. \textbf{Right panel}: average daily CSI standard deviation computed along the time dimension. For every sampled day, the standard deviation along the time dimension is computed for every pixel and then averaged over the 500 hundred days.} 
    \label{Avg_Std_plot}
\end{figure}

The CRPS serves as a compound metric, encompassing both reliability and sharpness to offer a holistic evaluation of model performance. It quantifies the distance between the ensemble and the optimal cumulative distribution for each pixel. This metric is then averaged across the entire test set (All-sky) and separately for low- and high-variability subsets, where variability is measured by the standard deviation computed on the input CSI fields. In the upper panel of \Cref{CRPS}, CRPS values are averaged across the test set for each pixel within the three test regions. Interestingly, similar spatial patterns are present in the right panel in Extended Data Figure \ref{Avg_Std_plot}, indicating a relation between standard deviation (variability) and CRPS values for the three models. In low-variability areas (Alps region in Extended Data Figure \ref{Avg_Std_plot}), the models, especially SHADECast and LDCast, exhibit a low CRPS. Conversely, the lower panel displays aggregated CRPS values averaged over all pixels, presenting the average, 25th, and 75th percentiles for each lead time.

SHADECast exhibits a 15\% improvement in overall CRPS compared to SolarSTEPS and a 7\% improvement over LDCast. A 120-minute SHADECast forecast is, then, as skillful as a 96-minute and 106-minute forecasts of SolarSTEPS and LDCast, respectively (see \Cref{CRPS}). This improvement, particularly evident in high-variability situations, suggests superior modeling of cloudiness evolution by SHADECast. The substantial enhancement over SolarSTEPS is attributed to differences in their CSI field generation mechanisms. SolarSTEPS simulates cloud evolution by random perturbations, generating CSI fields that share the spatial structure of the input satellite CSI maps but lack spatiotemporal information. In contrast, SHADECast models the spatiotemporal distribution of CSI maps, capturing information on spatial structure and temporal dynamics. The hypothesis is further supported by the smaller improvement in the low-variability subset, where cloudiness evolution is more static, resulting in similar performance between SolarSTEPS and SHADECast. This analysis underscores the importance of considering both, variability levels and the underlying dynamics of cloud evolution when assessing the efficacy of probabilistic forecasting models.

In low-variability situations, we notice a significant improvement of SHADECast over LDCast measured by an average improvement of $\sim 15\%$ in terms of CRPS. We attribute this finding to the conditioning nowcaster in SHADECast, which can better direct the forecast in low-variability situations, where a deterministic forecast contains more information with respect to a high-variability weather scenario. In these situations, the high-uncertainty regions are scarcer, so we expect the SHADECast ensemble to be closer to the nowcaster's forecast. 

\section*{Conclusion}

We have introduced a novel method for probabilistically forecasting SSI satellite maps that significantly outperforms existing approaches across diverse weather situations, from low to high variability scenarios. Our model stands out as the first ensemble-based approach capable of forecasting SSI satellite maps while adapting to dynamic weather conditions without suffering from blurriness and without requiring additional information beyond the input CSI fields.

Our model exhibits superior performance, consistently outperforming benchmarks (15\% and 7\% over SolarSTEPS and LDCast) across diverse weather situations, from low to high variability scenarios. 
This increased reliability is attributed to the incorporation of a deterministic latent nowcaster, which conditions the ensemble generation process. The modularity of our approach not only improves the performance but also permits the incorporation of alternative deterministic forecasting algorithm in our framework.

Built upon AFNO blocks and leveraging insights into cloudiness dynamics, SHADECast tackles the forecasting challenge by dividing it into a deterministic and a probabilistic components. The deterministic nowcaster forecasts low-uncertainty large-scale dynamics, whereas the probabilistic aspect is managed by the diffusion model, responsible for simulating the stochastic evolution of cloudiness fields at smaller scales. In this way, the generated ensemble can simulate the spatial structure and dynamics of cloudiness, enabling the prediction of extreme values.

Our contribution extends beyond theoretical advances, as SHADECast provides grid and trading operators with accurate and reliable forecast ensembles. This empowers them to enhance the integration of photovoltaic energy into the grid, mitigating the volatility impact on grid resilience. 

In conclusion, our model not only introduces a novel approach to SSI forecasting but also establishes a new standard in reliability and performance. By addressing the challenges of dynamic weather conditions and providing enhanced forecast ensembles, SHADECast contributes significantly to the advancement of energy meteorology and renewable energy integration.


\section*{Methods}
\subsection*{Solar Irradiance Dataset}
The clear-sky index (CSI) fields employed for this study are derived from spectral measurements of Earth taken by the Spinning Enhanced Visible and InfraRed Imager (SEVIRI) on board the Meteosat Second Generation geostationary satellite \cite{Schmetz2002}. The raw satellite images are processed by the HelioMont radiative transfer algorithm \cite{CASTELLI2014} to produce two-dimensional CSI fields. We refer to \cite{CARPENTIERI2023} for a comprehensive review of the dataset. The dataset spans 10 years from 2007 to 2016 at a temporal resolution of 15 minutes. The time period is motivated by constraints on data availability. The HelioMont CSI fields are only available for solar zenith angle (SZA) lower than 88\textdegree. The spatial resolution is approximately $0.02^{\circ} \times 0.02^{\circ}$. The region covered ranges from $8.3^{\circ}$E, $44.8^{\circ}$N to $12.8^{\circ}$E, $49.1^{\circ}$N corresponding to images of size $384\text{px}\times768\text{px}$ in the native Geostationary projection as shown in \Cref{GenerationProcess}. Missing pixels are filled by a linear three-dimensional (time, longitude and latitude) interpolation if they cover less than 2\% of the image, otherwise the image is discarded. 

\begin{figure}[H]    
    \centering
    \includegraphics[width=.7\textwidth, trim = 1.6cm 2.5cm 1.2cm 3cm, clip]{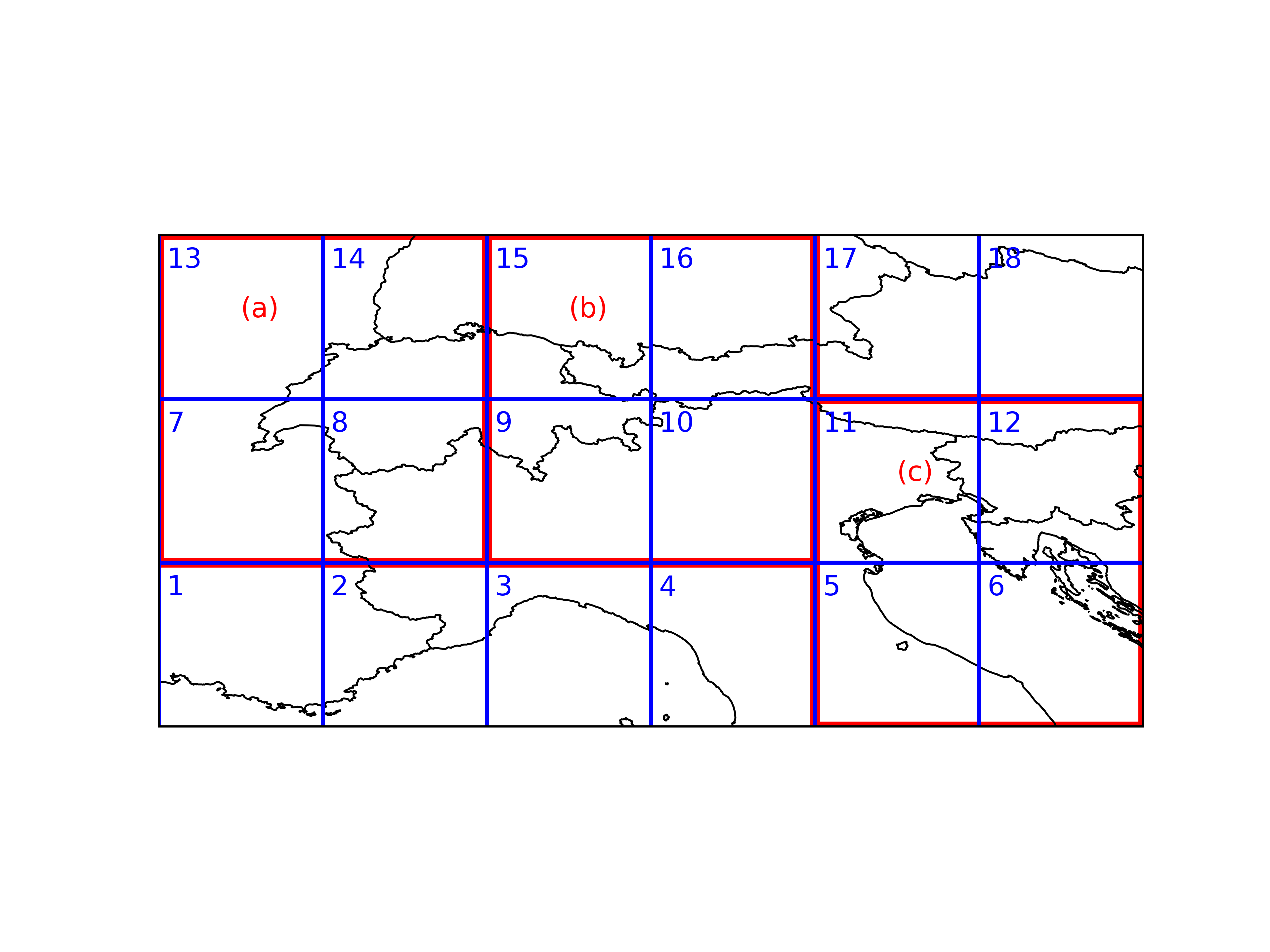} 
    \caption{Area covered by the HelioMont dataset \protect\cite{CASTELLI2014} The patches outlined in blue define the cropping applied to create the training set. For the test set we used three $256 \times 256$ patches identified by the red borders: (a), (b) and (c).} 
    \label{Patches}
\end{figure}

\begin{figure}[H]
    \centering
    \includegraphics[width=.8\textwidth, trim = 0cm 1cm 2.3cm 0cm, clip]{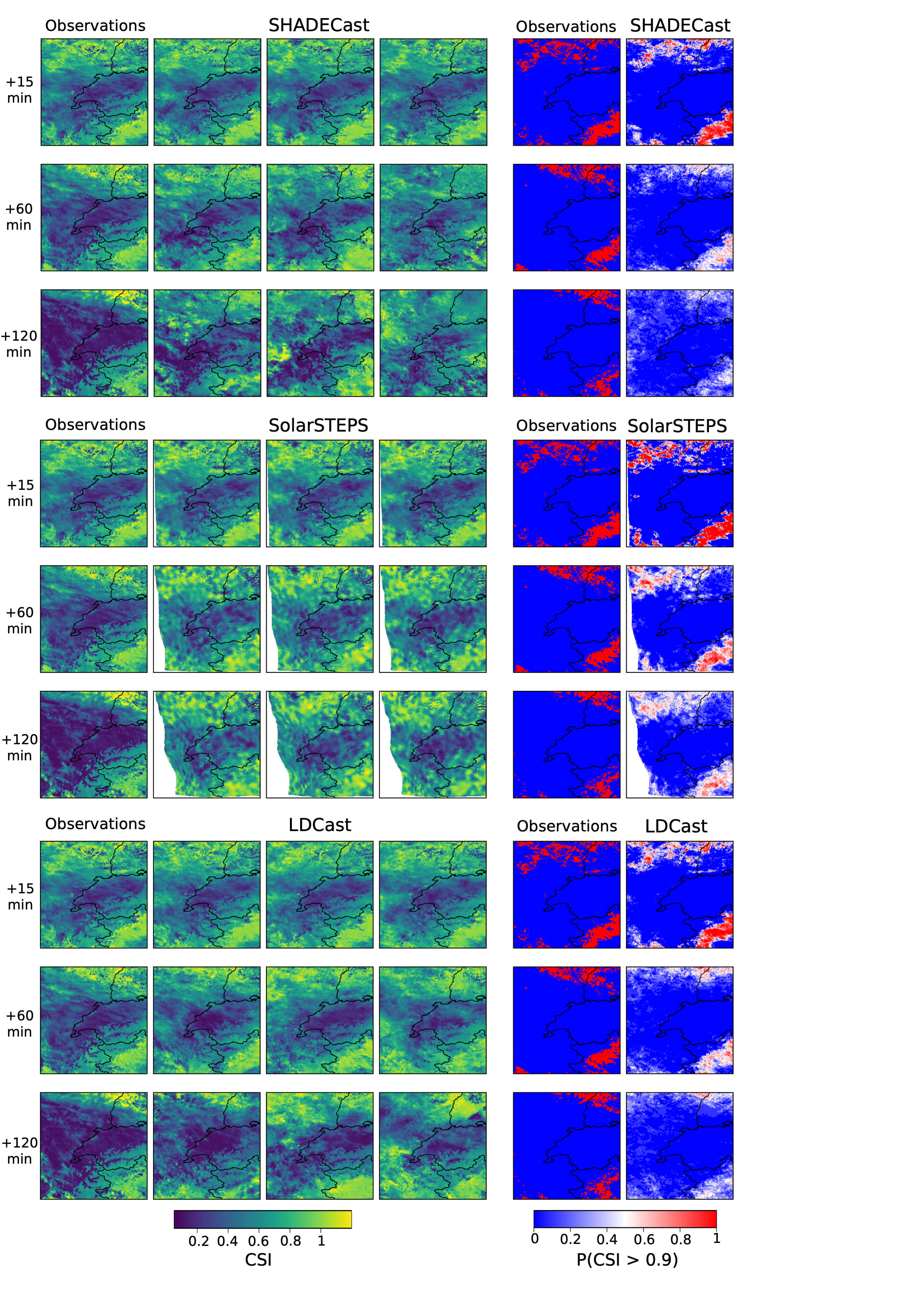} 
    \caption{Visualization of generated ensembles at three lead times for SHADECast and two benchmark models. The date (24 Feb. 2016) and starting time (11.45 am) are chosen to show a changing weather situation in which the cloudy surface (blue pixels) increases through the forecast. On the first column, the satellite CSI images are shown (Observations). The second, third and fourth columns show the best, average and worse ensemble members, respectively. The ensemble members are evaluated by their average RMSE over the entire forecast. The last two columns show the ground truth and forecasted probabilities of CSI exceeding $0.9$ (clear-sky). The forecasted region is patch (a) (see Extended Data \Cref{Patches}).} 
    \label{Test1_Forex}
\end{figure}

\begin{figure}[H]
    \centering
    \includegraphics[width=.7\textwidth, trim = 0cm 1cm 2.3cm 0cm, clip]{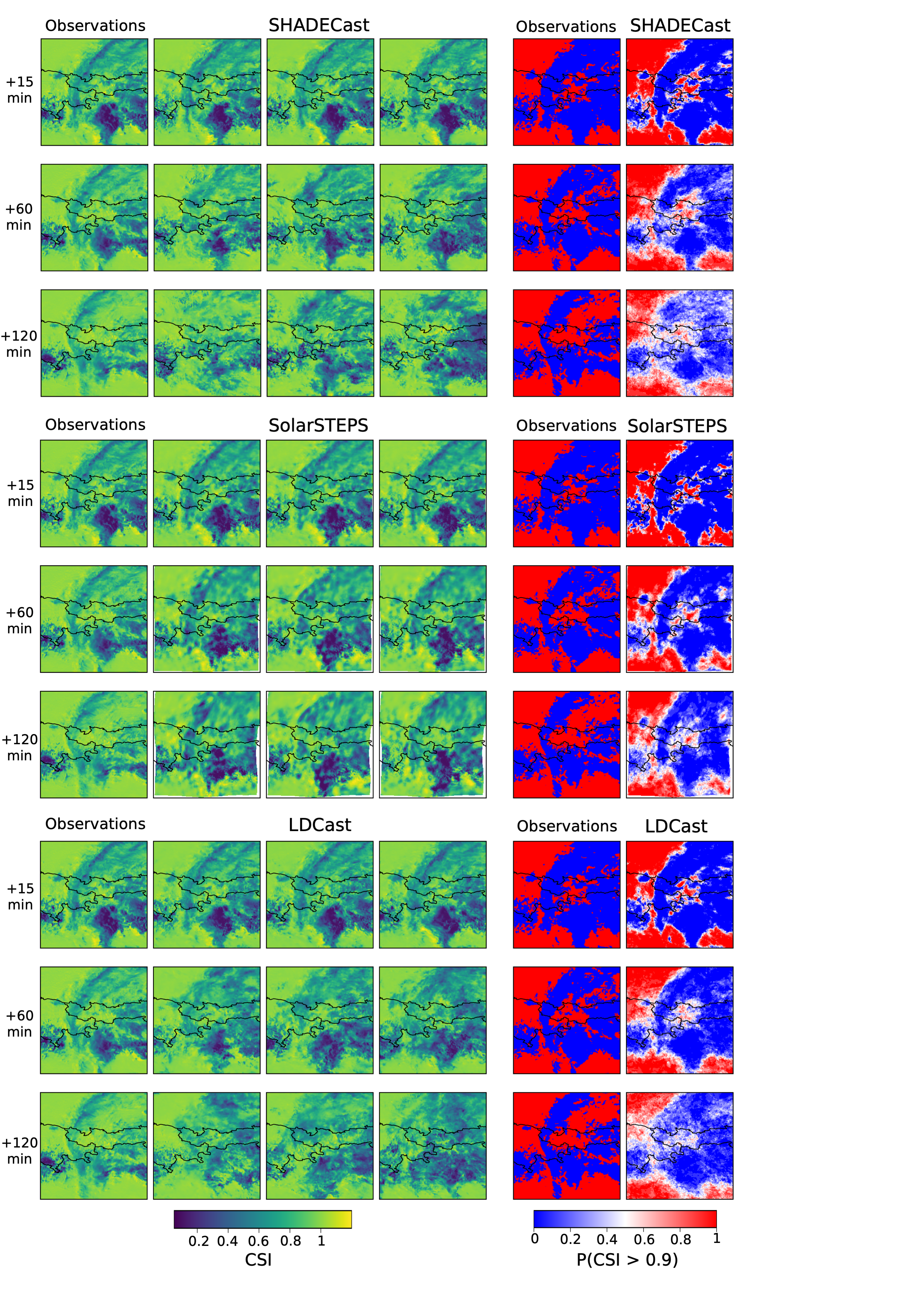} 
    \caption{Visualization of generated ensembles at three lead times for SHADECast and two benchmark models. The date (24 Jul. 2015) and starting time (05.30 am) are chosen to show a dissipation
    example of clouds on the bottom of the region. The figure is structured as Extended Data \Cref{Test1_Forex}. The forecasted region is patch (b) (see Extended Data \Cref{Patches}).} 
    \label{Test3_Forex}
\end{figure}

\begin{figure}[H]
    \centering
    \includegraphics[width=.7\textwidth, trim = 0cm 1cm 2.3cm 0cm, clip]{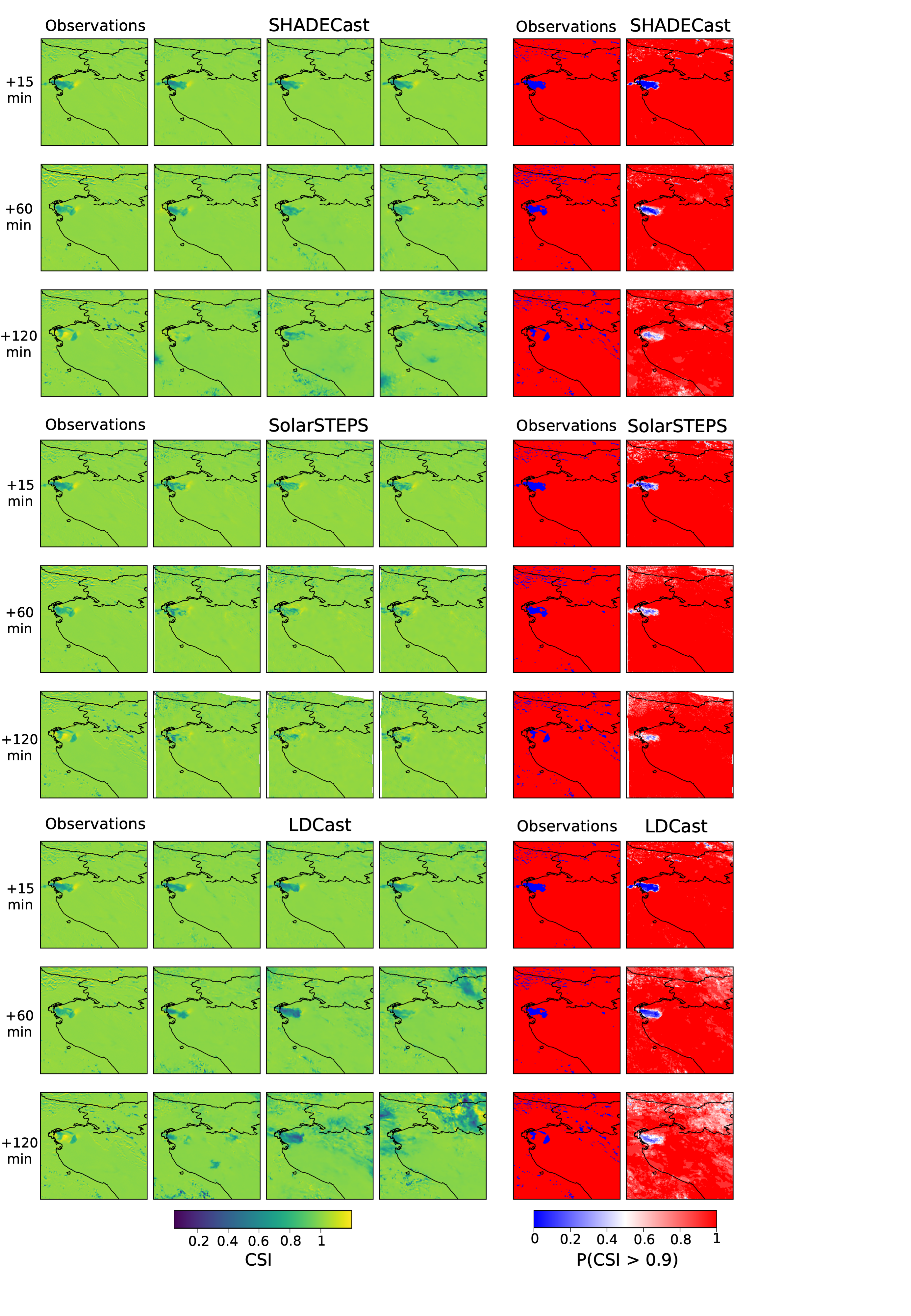} 
    \caption{Visualization of generated ensembles at three lead times for SHADECast and two benchmark models. The date (19 Mar. 2016) and starting time (11.15 am) are chosen to show a clear-sky and low variability weather example. The figure is structured as Extended Data \Cref{Test1_Forex}. The forecasted region is patch (c) (see Extended Data \Cref{Patches}).} 
    \label{Test5_Forex}
\end{figure}

Seven years of data are used for the model training (2007--2013) and one for the validation (2014), while two years are kept for the final testing (2015--2016). 
For training and validation, we cropped the maps into 18 $128\text{px}\times128\text{px}$ patches as shown in \cref{Patches}. Therefore, for the training set we have 18 regions and $365\times7$ days of data split into overlapping 12-step sequences (4 input and 8 output maps).

To create the test set, we randomly sampled 200 days from the 2-year period (2015--2016), and then randomly sampled 4 input sequences from each of the 200 days, resulting in 800 CSI satellite image sequences for every test set region. We make use of 3 $256\text{px}\times256\text{px}$ regions, namely the areas corresponding to patches (a), (b), and (c) as illustrated in \Cref{Patches}. Using larger images for the validation (with respect to the training set) accounts for the advection effect during the forecast lead time, aiming to reduce areas completely generated by the model. At maximum speed, clouds can cross most of the 128 pixels ($\approx250$ km) in less than 2 hours and so, the model would generate most of the forecast with no information on coming clouds. The use of smaller image patches in training was driven by memory and computational constraints. Notably, our model's architecture enables the forecast of arbitrarily large images. 

In Extended Data Figure \ref{Avg_Std_plot} we show average and standard deviation of CSI for every pixel covered by HelioMont dataset. The values are computed daily for 500 randomly sampled days from the training set and then averaged. 

\subsection*{SHADECast}
SHADECast is a conditional latent diffusion model incorporating Adaptive Fourier Neural Operator (AFNO) blocks \cite{guibas_afno_2022}, known for their efficacy in modeling chaotic systems like weather \cite{pathak2022}. With respect to current SSI forecasting models and LDCast \cite{leinonen2023}, the architectural innovation of SHADECast is the incorporation of an independently-trained AFNO-based forecasting model as conditioning model (nowcaster in \Cref{GenerationProcess}). The nowcaster focuses on forecasting large-scale components of the dynamics of cloudiness, while the diffusion model (denoiser) is responsible for forecasting the chaotic dynamics of small scales, thus generating ensembles of possible future evolutions.

The core concept of diffusion models entails forward diffusion and backward denoising processes\cite{Ho2020},\cite{Sohl-Dickstein2015}. The forward diffusion process iteratively introduces disruptive Gaussian noise into training data samples, whereas the backward process iteratively removes the noise from the noisy output of the forward process, restoring the data sample to its original state. Fundamentally, the denoising process is implemented to enable the model to learn the mapping of a known simple distribution (usually an uncorrelated Gaussian) to the data distribution, enabling the generation of realistic and accurate data samples.

\begin{figure}[H]
    \centering
    \includegraphics[width=\textwidth]{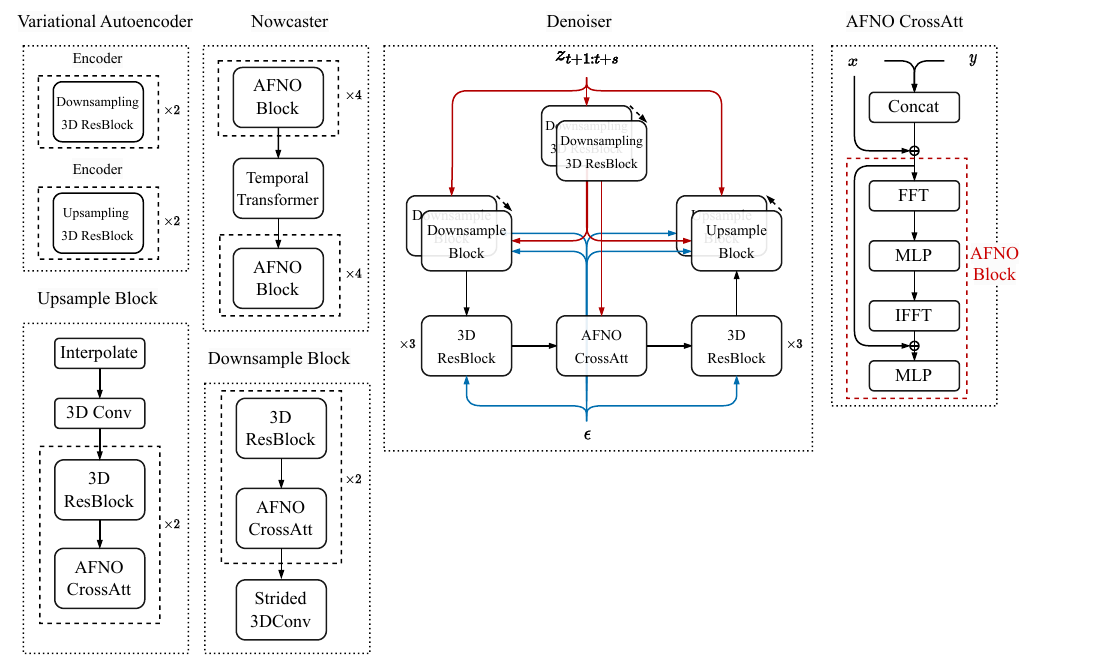} 
    \caption{A AFNO-based U-Net architecture is employed in our denoiser, alongside principal blocks integrated into the SHADECast architecture. The symmetrical design of the denoiser includes two downsampling and upsampling blocks. The latent forecast $z_{t+1:t+s}$ undergoes 3-dimensional strided residual blocks to match spatial dimensions with U-Net components, followed by concatenation with the output of AFNO cross attention blocks. In the right panel, $x$ represents the input from the previous layer, and $y$ is the conditioning input. Downsampling is achieved through strided 3D convolutional layers, whereas upsampling utilizes spatial axis interpolation. The 3-dimensional residual blocks consist of two convolutional layers connected by a skip connection.}
    \label{fig:Architecture}
\end{figure}

Our conditional latent diffusion model consists of three main components as depicted in Extended Data Figure \ref{fig:Architecture}:
\begin{enumerate}
    \item A \emph{variational autoencoder} (VAE), which compresses (decompresses) the data into (from) the latent space. Following the approach in \cite{Rombach2022}, modeling diffusion in the latent space achieves an optimal trade-off between accuracy and efficiency. 
    \item A latent AFNO-based deterministic \emph{nowcaster}. It takes the latent representation of the input CSI maps and forecasts consecutive maps in the latent space. The number of latent time steps is increased using a temporal transformer \cite{attention_is_all_you_need}. It can be used as an independent forecasting model.
    \item A latent \emph{denoiser}, which maps Gaussian noise to the future CSI maps in the latent space. Based on a U-Net architecture \cite{UNet}, it is conditioned on the nowcaster's output through AFNO Cross Attention blocks (Extended Data Figure \ref{fig:Architecture}). 
\end{enumerate}

The forecast generation process shown in \Cref{GenerationProcess} involves the encoding of $m$ past input CSI fields $C_{t-m+1:t}$ into the latent space, resulting in the latent tensor $z_{t}$ with an overall compression factor of 2. Then, the nowcaster performs a forecast in the latent space ($z_{t+1:t+s}$). $s$ represents the number of forecasted steps in the latent space, which are related to $n$ by $s=\frac{n}{c_t}$, where $c_t$ is the compression factor along the time dimension. Then, $z_{t+1:t+s}$ is employed to condition the denoiser that generates the forecast ensemble. The conditioning is performed by downsampling the deterministic forecast to match the dimensions of the U-Net layers of the denoiser (Extended Data Figure \ref{fig:Architecture}). The conditioning step is essential to guide the denoising process towards realistic future scenarios of cloudiness evolution. In summary, the goal of the denoiser is to project the input noise tensor ($\epsilon$) to the latent representation of the future $n$ satellite observations ($C_{t+1:t+n}$). It does so by iteratively performing numerous denoising steps \cite{Ho2020}. In our case, the CSI field sequence generation is governed by a pseudo-linear multistep sampler (PLMS) \cite{liu_plms_2022} to reduce the number of required denoising steps. PLMS permits to decrease the number of steps from $\approx1000$ to 25, maintaining the sample quality (refer to Supplementary Table \ref{PLMS_Steps_Val_CRPS}). Finally, the sampled sequence $\hat{z}_{t+1:t+s}$ is decoded back by the decoder to the final forecast ensemble member $\hat{C}_{t+1:t+n}$. 

\subsubsection*{Variational Autoencoder}
\label{VAE}
The variational autoencoder (VAE) exhibits a symmetrical architecture, as in \cite{kingma2022autoencoding}. The VAE's encoder processes 4-dimensional inputs, specifically sequences of CSI fields. This encoding phase consists of two downsampling 3-dimensional residual blocks, outputting two tensors, $\mu$ and $\Sigma$, namely the mean and covariance matrices of a Gaussian distribution. They serve as the foundation for the decoder's sampling process that samples a latent vector from the latent Gaussian and reconstructs it into a sequence of CSI fields. The downsampling and upsampling blocks in the VAE mirror those in Extended Data Figure \ref{fig:Architecture}, with the exception of the cross-attention layer.

The CSI field sequence is represented as a four-dimensional tensor with dimensions $(C, T, W, H)$, where $C$ denotes the number of image channels, $T$ is the time dimension, and $(W, H)$ represents the width and height of a single CSI map. In the latent space, dimensions $T$, $W$, and $H$ are reduced by a factor of 4, while $C$ is increased by a factor of 32, resulting in an overall compression factor of 2.

Regularization in the latent space is achieved through the Kullback--Leibler (KL) divergence between the latent data distribution and $N(0, 1)$. The reconstruction loss is quantified by the mean absolute error that measures the disparity between the VAE's input and the decoder output. The final loss is an interpolation between the two losses, with a coefficient of 0.05 for the KL loss. 

The VAE comprises approximately 800,000 parameters. For detailed architecture parameters, please refer to the training configuration file available in our GitHub repository.

\subsubsection*{Nowcaster}
\label{Nowcaster}
The AFNO-based nowcaster consists of four AFNO blocks, a temporal transformer \cite{attention_is_all_you_need}, and another four AFNO blocks. The AFNO blocks \cite{guibas_afno_2022} (Extended Data Figure \ref{fig:Architecture}) transform the input using a 3-dimensional Fast Fourier Transform (FFT) applied to the temporal and spatial axes. Subsequently, a multilayer perceptron (MLP) processes the transformed data along the channel dimension. Finally, the data undergoes inverse-FFT (IFFT), is summed with the original input, and processed by another MLP.

The temporal transformer is employed to increase the time steps through cross attention between the input and a sinusoidal time embedding tensor\cite{leinonen2023}. The time steps are increased by a factor of 2, resulting in $s=2$ in \Cref{GenerationProcess}. 

The nowcaster operates in the latent space following the approach in \cite{leinonen2023}. Computing the AFNO in the latent space aligns with the method in \cite{pathak2022}, where the authors utilized an embedding procedure to increase the channel dimension at the expense of $H$ and $W$. Through channel mixing in the Fourier space, we approximate global attention \cite{guibas_afno_2022}, as each pixel in the Fourier space contains information on the entire image.

The loss chosen is the Mean Absolute Error (MAE) and it is computed in the latent space. By computing the loss in the latent space we noticed two major improvements. First, we save one iteration (the decoding). Second, the forecasts result more detailed and less blurry even at longer lead times.

Overall, the architecture of the nowcaster comprises $\sim 6$M parameters. 
 
\subsubsection*{Denoiser}
\label{Denoiser}
The denoiser's AFNO-based U-Net architecture, depicted in Extended Data Figure \ref{fig:Architecture}, is symmetrical and comprises two main components: downsampling and upsampling blocks. The denoising process begins with the latent forecast $z_{t+1:t+s}$, which is downsampled with 3-dimensional strided residual blocks. This step is crucial for achieving spatial dimension alignment with the U-Net's downsampling and upsampling blocks. The resulting output is then concatenated with the output of AFNO cross attention blocks, denoted as $x$ and $y$ for the input from the previous layer and the conditioning input, respectively.

For downsampling, we employ strided 3D convolutional layers, effectively reducing spatial dimensions (height and width). Conversely, upsampling is achieved through interpolation on the height and width axis of the tensor. The backbone of the architecture consists of 3-dimensional residual blocks, featuring two convolutional layers with a skip connection to enhance feature extraction. The denoiser is trained to predict the noise as done in \cite{Rombach2022} and the chosen loss is the mean squared error (MSE). Moreover, as also done in \cite{Rombach2022}, the exponential moving average method is employed to stabilize the training.

This detailed architecture is visually represented in Extended Data Figure \ref{fig:Architecture}. The denoiser is defined by approximately $320$M parameters.

\subsection*{Data Processing and Training}
\label{Data Processing}
The number of CSI fields from Meteosat SEVIRI in a day depends on the daylight hours, resulting in a higher number of available data samples during summer as HelioMont cannot derive CSI at night. To mitigate this bias in our model, we generate each training sample by randomly selecting one day from the $365 \times 7$ available days and then selecting a sequence of maps from that day. This ensures that the models are trained on a balanced dataset, exposing them to an equal number of summer and winter sequences.

The validation set follows a similar sampling approach as the training set but with a fixed structure: sequences and days are sampled once, and these validation samples remain constant throughout the validation process. This is done to obtain a consistent validation through the training process.

To facilitate model convergence and performance, the data are normalized by mapping the values to the $\bigl[-1, 1\bigr]$ range. The normalization is straightforward as HelioMont CSI values are bounded in the $[0.05, 1.2]$ range.

The three components of SHADECast (autoencoder, nowcaster and denoiser) are trained independently. The training and validation sets are the same for the three training processes. The training is terminated if the validation loss does not decrease for at least 10 epochs (early stopping). Moreover, after 5 epochs without improvement, the learning rate is divided by a factor of 4. The initial learning rate for VAE and the nowcaster is $10^-3$, while for the Denoiser we set $10^-4$. Similarly, the batch size is set to 256, 240 and 96 for the VAE, Nowcaster and Denoiser, respectively. The batch sizes are chosen to maximally exploit the available GPU memory and stabilize the training.

\subsubsection*{Computational requirements}
The training of the SHADECast diffusion model requires approximately 500 GPU hours on 24 Nvidia P100 GPUs. 
Generating a 2-hour forecast ensemble of 10 members at 15-min resolution for $256\times256$ pixels images, takes 1 minute on a single Nvidia P100 GPU.

\subsection*{Benchmark models}
\label{Benchmark models}
To assess our model's performance, we compare it to two probabilistic ensemble-based benchmark models: SolarSTEPS \cite{CARPENTIERI2023_SS} and LDCast \cite{leinonen2023}. In order to use the latter, we adapted and trained the original precipitation nowcasting model to forecast cloudiness. It shares the same architecture as SHADECast (see Extended Data Figure \ref{fig:Architecture}) for a fair comparison. In LDCast, the forecaster component is trained together with the denoiser and is, in fact, a feature extractor on the input maps. Therefore, the model is not conditioned on a deterministic forecast but, indirectly, on the input CSI fields. LDCast is chosen to illustrate that our physics-motivated choice of a separate nowcaster indeed improves the forecast accuracy and reliability of the ensemble. Overall, the training procedure and the data used are the same as for SHDECast training.

SolarSTEPS \cite{CARPENTIERI2023_SS} is an optical-flow based approach, which was shown to outperform state-of-the-art models in the task of probabilistically forecasting satellite-derived CSI maps over Switzerland. Therefore, we consider it a valuable benchmark case in the present paper. The SolarSTEPS approach is based on the scale-dependent temporal variability of cloudiness: the small scales have a shorter lifetime with respect to bigger scale. The different scales' temporal evolution is thus modeled independently by different linear AR models. The approach permits the model to predict both the motion (optical-flow) and evolution (AR models) of cloudiness. The ensemble generation is governed by perturbing the AR models with a novel technique to generate spatially correlated CSI fields based on the short-space Fourier transform \cite{Nerini_ssft_2017}. The method presented in \cite{Nerini_ssft_2017} is modified to take into account the variability of the input maps in the generation of the perturbing fields. Moreover, SolarSTEPS has shown to outperform trivial benchmark models such as the persistence model. The parameterization used in our evaluation reflects the one presented in \cite{CARPENTIERI2023_SS}.

Moreover, we investigate the advantages of our probabilistic approach in comparison to the state-of-the-art deterministic model in cloudiness forecasting, IrradianceNet \cite{nielsen_irradiancenet_2021}. The model architecture remains consistent with the original paper. We retrained the model using only cloudiness fields on our training set as done for LDCast and SHADECast. In the original paper, the authors conducted a 2-step forecast. For comparability with other models, we autoregressively forecast 8 steps into the future. The model is trained on $128\times128$ images and tested on $256\times256$ similarly as done in \cite{nielsen_irradiancenet_2021}. Due to the model architecture limitations, forecasting arbitrarily large images is not possible. Consequently, a linear interpolation is applied on the borders of the individual forecasts, as detailed in \cite{nielsen_irradiancenet_2021}. It is important to note that the output interpolation introduces visible artifacts along the borders of single forecasts.

\subsection*{Performance Metrics}
\label{Metrics}
The evaluation of the forecast ensembles is carried out by using probabilistic and deterministic metrics. For probabilistic forecasts, the main properties we evaluate are the reliability and sharpness of the forecast ensembles \cite{gneiting_2007}. 

A reliable forecast ensemble is characterized by the observed value falling within the predicted ensemble. In an ideal scenario where the model accurately captures the uncertainty of the dynamics, the observations should be uniformly distributed within the ensemble. To assess this distribution, rank histograms \cite{Rank_histograms} depict the frequency of the observed value's location among the ensemble members. In practical terms, a concave histogram signals under-confidence, indicating that the model tends to overestimate uncertainty. This results in forecasts with excessively high variance, suggesting a wider range of possibilities than observed. Conversely, a convex histogram signals overconfidence, indicating that the ensemble is too narrow and fails to adequately capture the actual uncertainty in the system dynamics. In such cases, the forecasted range is too restrictive, leading to potential underestimation of the true variability in the observed values.

Reliability is also described by the Prediction Interval Coverage Probability (PICP). PICP measures the percentage of observed values that lie in the ensemble prediction interval. We randomly sample 1000 pixels for each lead time image and check whether they fall inside the 5\% and 95\% percentiles of our forecast. However, PICP does not provide any information on the forecast informativeness, as an overdispersive model could lead to high PICP values. For this reason, we also measure the Prediction Interval Normalized Averaged Width (PINAW). PINAW measures the width of the prediction interval and so, it provides information on the forecast sharpness. An ideal forecast should reflect high PICP values and a low PINAW. 

The Continuous Ranked Probability Score (CRPS) is employed to evaluate the overall quality of probabilistic SSR forecasts \cite{gneiting_2007}, \cite{CRPS}. CRPS accounts for both reliability and sharpness. It does so by measuring the distance between the cumulative density function of the ensemble $F$ and the Heaviside function $H$ centered on the observation $y$. The normalized CRPS for the $i$-th pixel is then defined as:
\begin{equation}
    \textrm{nCRPS}_i = \frac{1}{\textrm{CSI}_{\textrm{max}}} \int_{-\infty}^{+\infty} (F_i(c)-H(c-y_i))^2 dc
\end{equation}
The Heaviside function centered in $y_i$ represents the ideal cumulative distribution for a perfect probabilistic forecast and $F_i$ is the forecasted cumulative distribution for the $i$-th pixel. CRPS, then, measures the distance between the Heaviside function and $F_i$ for every point $c$ in the $F_i$ domain. It is computed at pixel level and averaged for every forecast step, ending up with a CRPS value for every forecasted pixel. We consider the normalized CRPS (nCRPS) by normalizing the CRPS with the maximum clear-sky index value, which is $\textrm{CSI}_{\textrm{max}}=1.2$. 

Finally, the normalized Root Mean Square Error (nRMSE) is used to measure the accuracy of the ensemble mean. The ensemble mean serves as a representative estimate of the central tendency of the forecasted distribution. Evaluating its accuracy provides insights into how well the ensemble captures the expected or average outcome. The nRMSE for a forecasted map is defined as:
\begin{equation}
    \textrm{nRMSE} = \frac{1}{\textrm{CSI}_{\textrm{max}}} \sqrt{\frac{1}{N}\sum_{i=1}^{N}(\hat{y}_i-y_i)^2}
\end{equation}

\section*{Data Availability}
The HelioMont data are licensed and can be obtained from the MeteoSwiss customer service via \\ https://www.meteoswiss.admin.ch/home/form/customer-service.html.

\section*{Code Availability}
The code to train and test SHADECast is made available at: \\ https://github.com/AECML/GenerativeNowcasting. \\
The code for SolarSTEPS is made available at: https://github.com/AECML/SolarSTEPS. \\
The original LDCast model is available at: https://github.com/MeteoSwiss/ldcast, whereas in the SHADECast repository there is the architecture adapted to forecast cloudiness fields.

\section*{Acknowledgments}
We acknowledge funding from the Swiss National Science Foundation (grant 200021\_200654) and from the Swiss National Supercomputing Centre (CSCS) under project ID s1144. JL was supported by the fellowship ``Seamless Artificially Intelligent Thunderstorm Nowcasts'' from the European Organisation for the Exploitation of Meteorological Satellites (EUMETSAT). The hosting institution of this fellowship was MeteoSwiss in Switzerland.

\section*{Contributions}
A.C., D.F. and A.M. managed the project. A.C. and D.F. conceptualized the model. A.C. and J.L. conceptualized the software.
A.C. created data sets, wrote the software and conducted experiments. A.C., D.F., J.L. and A.M. wrote the paper. A.M. and D.F. managed funding, licensing, legal agreements and computing resources.

\extendeddata

\resetfigurelabel
\clearpage
\printbibliography[filter=onlymain, title={References}]

\begin{refsegment} 
\clearpage

\suppmat
\appendix

\section*{Supplementary Material}
\label{supplementary_material}

\subsection*{Clear-sky index extreme values}
Deterministic forecasting models such as \cite{nielsen_irradiancenet_2021} tend to produce blurry forecasts after few steps. The blurriness results in unrealistic forecasts, which do not respect the spatial structure of the ground truth cloudiness field. This is due to the training process of minimizing a pixel-level loss function, such as mean squared error or mean absolute error \cite{babaeizadeh2018stochastic}. In practice, this translates to forecasts converging towards a mean value, impeding the accurate forecast of extreme values. In our case, extreme values represent extreme overcast or complete clear-sky weather situations. We define extreme overcast and clear-sky with CSI values below 0.15 and over 0.95, respectively. 

To measure the ability of the forecasting models to predict such extremes, we make use of the Fraction Skill Score (FSS) metric \cite{fss, leinonen2023}. FSS evaluates the fraction of correctly predicted area for a specific threshold of interest, indicating how well a model captures the spatial distribution of an event. A higher FSS suggests better spatial agreement between predicted and observed phenomena.

In Supplementary Figure \ref{FSS}, FSS values are shown for SHADECast and the benchmark models for clear-sky and overcast situations. The metric is shown for the entire test set (All-Sky), low- and high-variability subsets. On average, the probabilistic models perform better than ConvLSTM due to their sharp forecasts, which do not suffer from increasing blurriness. In fact, ConvLSTM forecasts perform discretely well in low variability situations (central column) and on average in the first 15 to 30-minute. After few steps, the accuracy degrades. On the other hand, SHADECast outperforms the benchmark models on predicting CSI values higher than 0.95, especially in high variability situations with a 16\% improvement over ConvLSTM. Low variability situations usually are defined by the absence of clouds, so extreme low values are challenging to predict for all forecasting models (see second panel, central column in Supplementary Figure \ref{FSS}).  However, SHADECast results to outperform the other models, improving clear-sky ConvLSTM's FSS by 28\%. 

\begin{figure}[h]
    \centering
    \includegraphics[width=\textwidth, trim={6cm 0 1cm 0},clip]{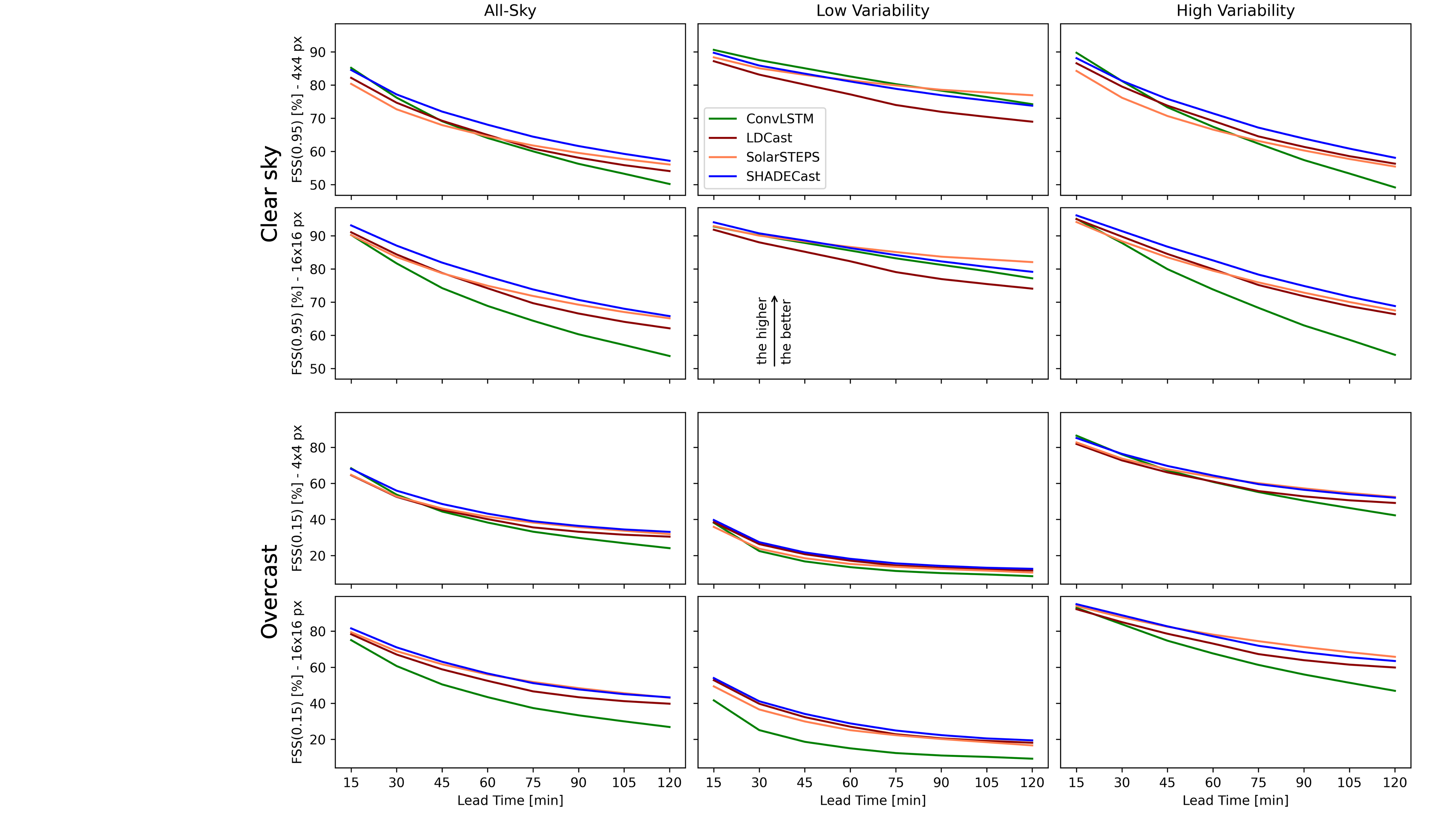} 
    \caption{Fraction skill score (FSS) with threshold set to 0.95 and 0.15 relative to two window sizes: $4\times4$ pixels and $16\times16$ pixels. For SHADECast, LDCast and SolarSTEPS, the FSS is computed for every ensemble member and then averaged. The procedure is then applied for the entire test set (All-sky) and two subsets.}
    \label{FSS}
\end{figure}

\subsection*{Forecast spatial structure}
To showcase the quality of our model, and more in general, of probabilistic modeling with respect to deterministic approaches, power spectra can be employed to measure the degree of similarity of a forecast to the ground truth \cite{Nerini_ssft_2017}. In Supplementary Figure \ref{All_PS_Test1}, power spectra are shown to demonstrate the effects of blurriness in deterministic forecasts. In fact, the convLSTM generated fields do not respect the spatial structure of CSI fields as the power spectrum gets further from the observation with time and increasing blurriness. On the other hand, ensemble-based models provide spatially consistent forecasts for scales up to 3 pixels through the entire forecast.

\begin{figure}[h]
    \centering
    \includegraphics[width=\textwidth]{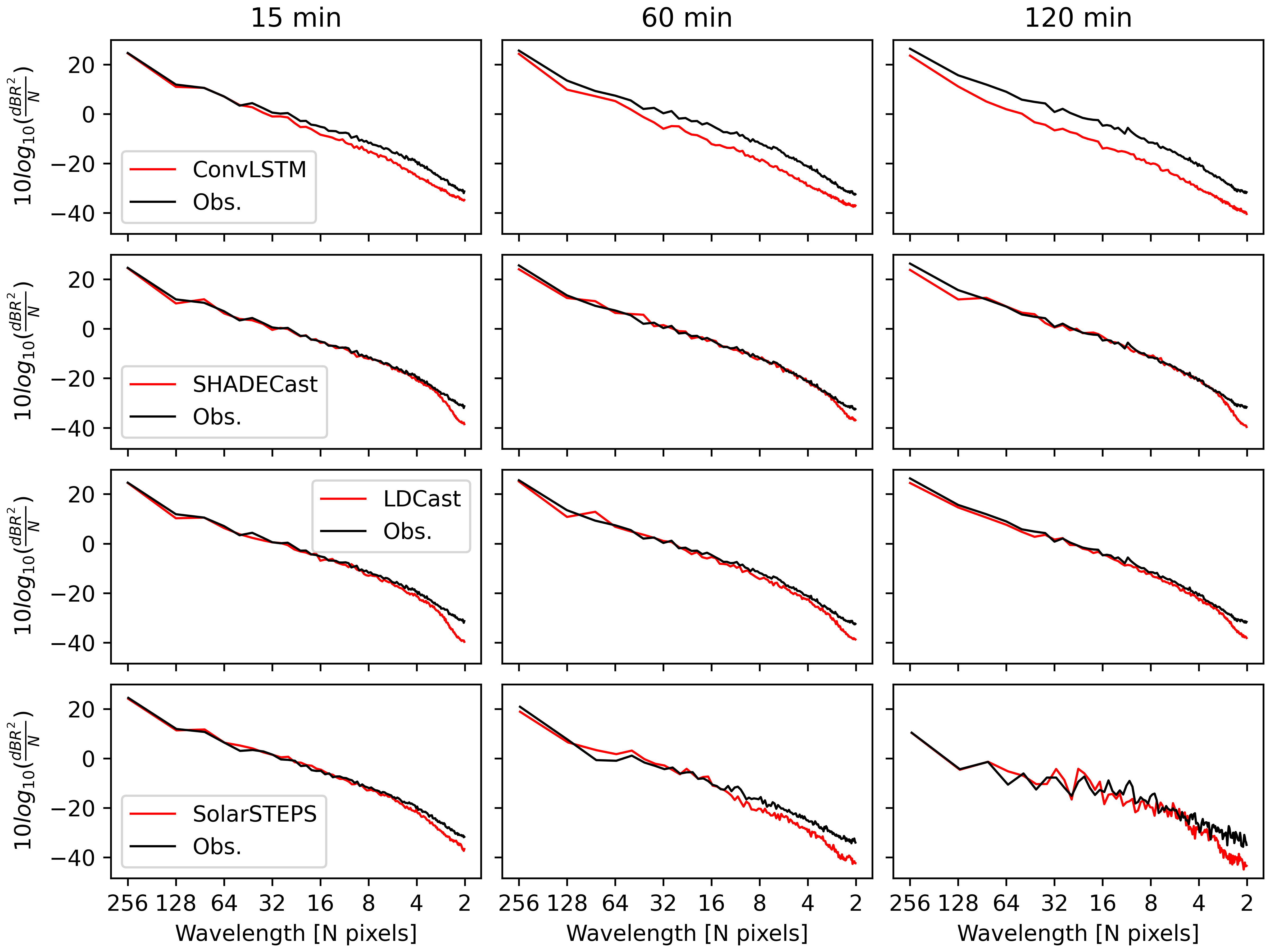} 
    \caption{One dimensional power spectra at three lead times relative to the case study shown in \Cref{GenerationProcess}. For SolarSTEPS, we did not count the missing values for both the forecasted fields and the ground truth. The probabilistic models clearly outperform the deterministic convLSTM \cite{nielsen_irradiancenet_2021} in terms of spatial structure of the forecast.}
    \label{All_PS_Test1}
\end{figure}

\subsection*{Denoising steps}
Diffusion models are trained to denoise single steps in the backward process \cite{Ho2020}. However, in the generation, they need to perform all the denoising steps to map Gaussian uncorrelated noise to the data distribution. The number of steps required is high, making the generation process expensive. The pseudo linear multi-step (PLMS \cite{liu_plms_2022}) algorithm solves the backward diffusion process employing only few steps. 

A grid search method is employed to find the optimal number of PLMS steps. We run SHADECast with 10, 25 and 50 PLMS steps on 200 sequences randomly sampled from the validation set (2014). Supplementary Table \ref{PLMS_Steps_Val_CRPS} shows the results of our grid search. 25 is the optimal number of steps as it performs better than 10 and similarly to 50 but with lower computational requirements.
\begin{table}[h]
\centering
\begin{tabular}{|c|ccc|ccc|ccc|}
\hline
PLMS Steps & \multicolumn{3}{c|}{10} & \multicolumn{3}{c|}{25} & \multicolumn{3}{c|}{50} \\ \hline
Lead Time {[}min{]} &
  \multicolumn{1}{c|}{+15} &
  \multicolumn{1}{c|}{+60} &
  +120 &
  \multicolumn{1}{c|}{+15} &
  \multicolumn{1}{c|}{+60} &
  +120 &
  \multicolumn{1}{c|}{+15} &
  \multicolumn{1}{c|}{+60} &
  +120 \\ \hline
All Sky &
  \multicolumn{1}{c|}{0.047} &
  \multicolumn{1}{c|}{0.078} &
  0.103 &
  \multicolumn{1}{c|}{0.045} &
  \multicolumn{1}{c|}{0.073} &
  0.096 &
  \multicolumn{1}{c|}{0.045} &
  \multicolumn{1}{c|}{0.074} &
  0.096 \\ \hline
Low Var. &
  \multicolumn{1}{c|}{0.030} &
  \multicolumn{1}{c|}{0.056} &
  0.080 &
  \multicolumn{1}{c|}{0.027} &
  \multicolumn{1}{c|}{0.050} &
  0.071 &
  \multicolumn{1}{c|}{0.028} &
  \multicolumn{1}{c|}{0.050} &
  0.073 \\ \hline
High Var. &
  \multicolumn{1}{c|}{0.063} &
  \multicolumn{1}{c|}{0.101} &
  0.127 &
  \multicolumn{1}{c|}{0.060} &
  \multicolumn{1}{c|}{0.097} &
  0.118 &
  \multicolumn{1}{c|}{0.061} &
  \multicolumn{1}{c|}{0.097} &
  0.120 \\ \hline
\end{tabular}

\caption{Validation set nCRPS for SHADECast with different numbers of PLMS denoising steps.} 
\label{PLMS_Steps_Val_CRPS}
\end{table}

\subsection*{Additional results}
Here we provide further details on the models performance. In Supplementary Figure \ref{All_Rank_Histograms}, we show the rank histograms for SHADECast and benchmarks for the three test regions (a, b, c), individually. SHADECast uniformly outperform the baseline models in every region. Similarly, in Supplementary Figure \ref{All_CRPS}, the CRPS is shown for the different regions, lead times and weather situations. Finally, in Supplementary Figure \ref{RMSE}, the accuracy of the ensemble mean is measured through the normalized RMSE. SHADECast provides the most unbiased ensembles compared to the benchmarks, further highlighting the modeling superiority of our approach.

\begin{figure}[h!]
    \centering
    \includegraphics[width=\textwidth]{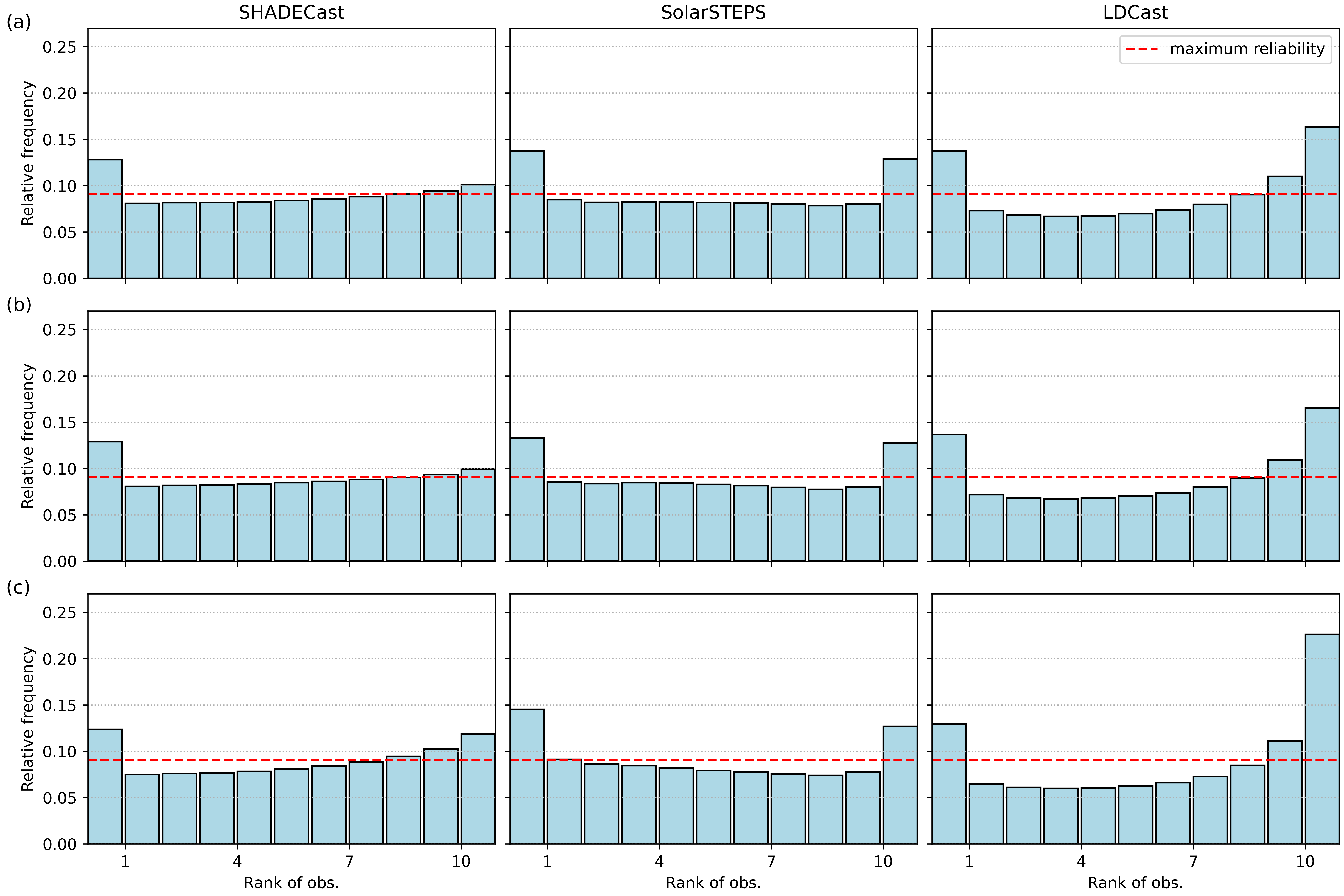}
    \caption{Rank histograms for the test set for the three locations considered. The panel names correspond to the test patches shown in Extended Data Figure \ref{Patches}. The models reliability shows similar patterns at the three test locations. However, SHADECast shows an higher reliability at locations (a) and (b).}
    \label{All_Rank_Histograms}
\end{figure}

\begin{figure}[h!]
    \centering
    \includegraphics[width=\textwidth]{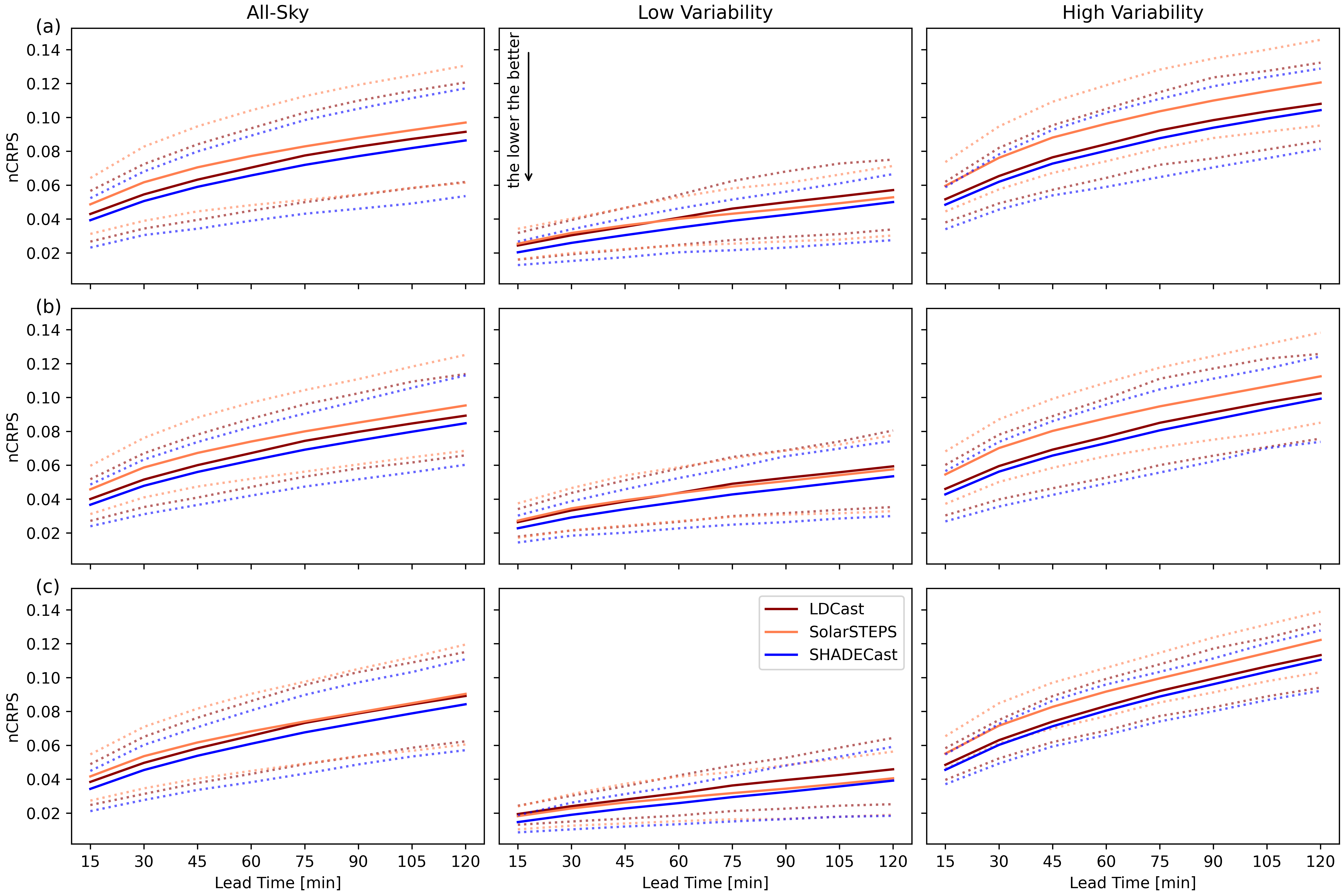}
    \caption{The CRPS plots shown are relative to the three different locations used in the test set. For each location we retrieved the low- and high-variability samples.}
    \label{All_CRPS}
\end{figure}

\begin{figure}[h!]
    \centering
    \includegraphics[width=\textwidth]{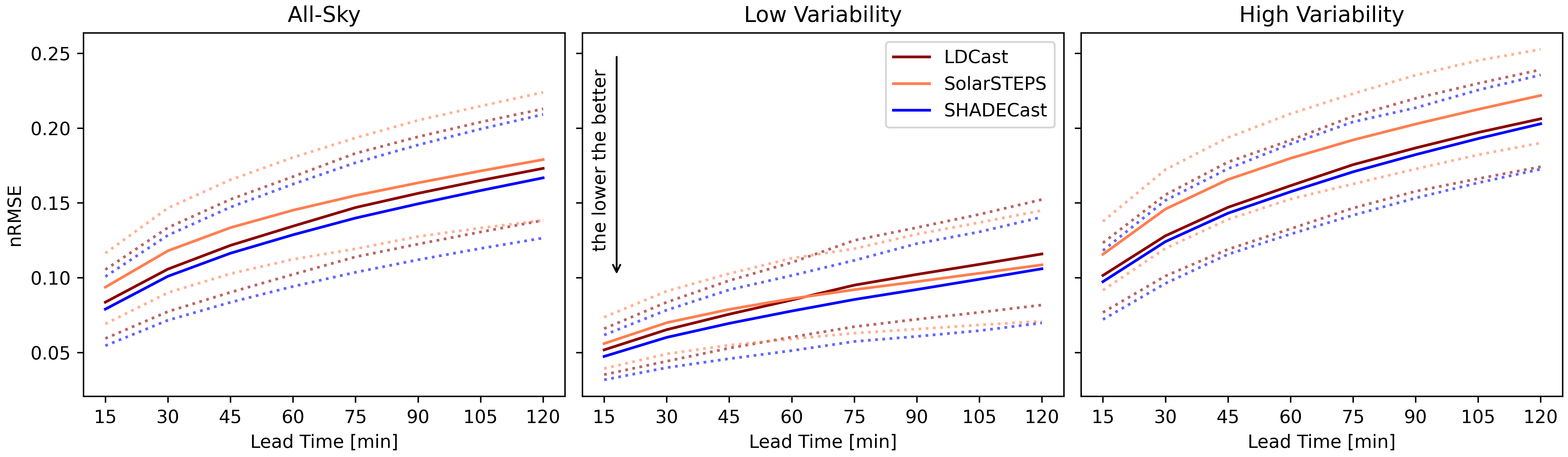}
    \caption{The root mean squared error is averaged among all the test samples for each lead time in the forecast. The plot shows the results for SHADECast and the two benchmarks for the entire test set (All-sky) and low- and high-variability samples. The dotted lines define the 25\% and 75\% percentiles.}
    \label{RMSE}
\end{figure}

\clearpage

\clearpage
\printbibliography[filter=onlyappendix, title={References}]
\end{refsegment}
\end{document}